\documentclass[preprint]{aastex}

\shorttitle{Mid-IR globular cluster colors}
\shortauthors{Barmby \&  Jalilian}

\begin{document}

\title{Comparing Mid-Infrared Globular Cluster Colors With Population Synthesis Models}

\author{P. Barmby and F. F. Jalilian}
\affil{Department of Physics \& Astronomy, University of Western Ontario, London, ON N6A 3K7, Canada}

\begin{abstract}
Several  population synthesis models now predict integrated colors of 
simple stellar populations in the mid-infrared bands. To date, the models have not
been extensively tested in this wavelength range. In a comparison of the  predictions of 
several recent  population synthesis models, the integrated colors are found to
cover approximately the same range but to disagree in detail, for example on
the effects of metallicity. To test against observational data, 
globular clusters are used as the closest objects to idealized
groups of stars with a single age and single metallicity.
Using recent mass estimates, we have compiled a sample of massive, old globular
clusters in M31 which contain enough stars  to guard against the stochastic  effects 
of small-number statistics, and measured their integrated colors in the \textit{Spitzer}/IRAC bands.
Comparison of the cluster photometry in the IRAC bands with the model predictions shows that the models reproduce the
cluster colors reasonably well, except for a small (not statistically significant)
offset in $[4.5]-[5.8]$. In this color, models without circumstellar dust emission predict bluer values
than are observed. Model predictions of colors formed from the $V$ band and the 
IRAC 3.6 and 4.5~\micron{} bands are redder than the observed data at high metallicities
and we discuss several possible explanations.
In agreement with model predictions, $V-[3.6]$ and $V-[4.5]$ colors 
are found to have  metallicity sensitivity similar to or slightly better than $V-K_s$.
\end{abstract}


\section{Introduction}

Star clusters are popular probes of the history of galaxy and star formation.
As compact systems with relatively homogeneous properties, their stellar
and dynamical histories can be modeled with reasonable accuracy, at least
compared to their parent galaxies. Studies of extragalactic star clusters have
experienced a major step forward with the availability  of the \textit{Hubble Space Telescope},
which provided the spatial resolution necessary to identify clusters in distant
galaxies and resolve those in nearby galaxies into individual stars,
and large ground-based telescopes, which provide the light-gathering power
to study these faint objects spectroscopically.
The study of extragalactic GCs has provided important clues to the
process of galaxy assembly and complements the direct study of high-redshift galaxies
\citep{brodie06}.

Star clusters also provide important tests of simple stellar population (SSP)  models 
constructed via population synthesis methods.
Although the evidence is now overwhelming that globular clusters are not
single-age, single-metallicity populations  
as previously believed \citep[see the review by][]{bragaglia10}, they
are still less complex than the population mix found in even the smallest galaxies.
As such, they provide important observational tests for population synthesis models, and
there have been many examples of such tests \citep{peacock11,riffel10,conroy10,pessev08,barmby00a,bica86a}.
Although becoming increasingly important in the study of nearby
stellar populations \citep[e.g.][]{sheth10}, and having the advantage of reduced
extinction sensitivity, the model predictions  at mid-infrared wavelengths $\lambda \gtrsim 3$~$\mu$m,
have not been well-tested.
Mid-infrared observations from ground-based telescopes are impractical for
all but the brightest sources, due to the high background, 
and the existing stellar libraries may not yet be adequate
\citep[although see][]{vandenbussche02,sloan03,hodge04}. 
The ATLAS9 model atmospheres, on which many population synthesis models are at least partly based,
have line opacity spectra calculated to a maximum wavelength of 10~\micron.
Modeling infrared stellar spectra  with a Rayleigh-Jeans law is quite common
\citep[e.g.][]{zasowski09}, but this is
 ``clearly a crude approximation'' \citep{marigo08}. Those authors state that  bolometric corrections
for the coolest stars are uncertain by up to a few tenths of a magnitude.
In their investigation on the use of model atmospheres for infrared calibration, \citet{decin07}
found that model spectra are accurate to 3\% in the mid-IR
but also noted that ``for the purpose of calculating theoretical spectra in the mid to far-IR, 
some line lists are still  far from complete or accurate''.

The \textit{Spitzer} Space Telescope probes these mid-infrared wavelengths.
While young star clusters have been the subject of numerous investigations
with \textit{Spitzer} \citep[e.g.][]{gutermuth09}, there has been relatively little analysis of  
the mid-infrared properties of old star clusters. 
Old star clusters' declining spectral energy distribution 
in the mid-infrared, in combination with the presumed lack of interesting spectral features,
is presumably the reason for this lack of attention.
Mid-infrared observations have been used in
studies of mass-losing stars in individual clusters \citep{boyer06,boyer08,boyer10,
origlia07,origlia10} as well as searches for cold dust in the intracluster
medium \citep{matsunaga08,barmby09}.
One of the few studies of  extragalactic globular cluster systems is that of \citet{spitler08}.  
The present paper provides mid-infrared photometry of  massive globular
clusters in the Andromeda galaxy, \object{M31}, taken with the \textit{Spitzer Space Telescope}, and compares 
them to population synthesis models which predict magnitudes in the IRAC bands, namely
GALEV \citep{kotulla09}, FSPS \citep{conroy09} and models from the Padova group \citep{marigo08}.
Only the four IRAC bands are used here.  M31 clusters are generally too faint to be detected in the MIPS
observations of \citet{gordon05}, and
photometry of the handful of Milky Way clusters observed with the MIPS instrument is published 
elsewhere \citep{boyer06, boyer08, barmby09}.

Throughout this work we use the Vega magnitude system and
assume a distance to M31 of 780~kpc \citep{mcconnachie05}, 
for which 1\arcsec\ corresponds to 3.78~pc. 

\section{Data and Models}

\subsection{Cluster Mass Limit}

Star clusters contain a limited number of stars and
photometry of star clusters must consider stochastic effects \citep{buzzoni89, renzini98}. The
integrated light of a cluster can be dominated by a few bright stars,
particularly in the infrared where the luminosity of the asymptotic
giant branch is highest \citep{ fouesneau10}. For this first comparison of star cluster
photometry with SSP models in the mid-infrared, we wanted to avoid stochastic effects as
much as possible, so we sought to select only the most massive clusters.
One criterion for choosing a selection limit is the lowest luminosity limit (LLL) defined
by \citet{cervino04}.
This limit is defined for a specific SSP model
as the luminosity of the brightest star included in the model, at the age,
metallicity, and wavelength under consideration.
For 10~Gyr old clusters, \citet{cervino04} found the total mass corresponding to the
LLL in the $K$-band to be approximately  $10^5$~M$_{\sun}$;
those authors showed that a total mass of about 10 times this limit is required 
in order for stochastic effects to  cause less than a 10\% dispersion in the total luminosity.

Related to the LLL approach is the concept of the ``effective number of stars'' emitting
at a given wavelength, introduced by \citet{buzzoni89}. With $N_{\rm eff}$ defined via
$\delta L_{\rm tot}/ L_{\rm tot} = N_{\rm eff}^{-1/2}$, the value of $N_{\rm eff}$ for a
10\% uncertainty in luminosity is $N_{\rm eff}=(0.1)^{-2}=100$. For a 10\% color uncertainty,
assuming the same $N_{\rm eff}$ for both bands, $N_{\rm eff}\geq 200$ \citep{buzzoni05}.
The Evolutionary Population Synthesis (EPS) website\footnote{\url{http://www.bo.astro.it/$\sim$eps/home.html}}
tabulates model predictions of $N_{\rm eff}(\lambda)$ as a function of age for populations with $L_{\rm bol} = 10^{10}L_{\sun}$.
For ages of 10--15 Gyr, typical values of $N_{\rm eff}(4\, \mu{\rm m})$ range from $10^{6.6}-10^{6.85}$ depending
on the metallicity, IMF and horizontal branch treatment;
for ages of 5 Gyr, $N_{\rm eff}(4\, \mu{\rm m})\sim10^{6.4}$. To scale  $N_{\rm eff}$ from $L_{\rm bol} = 10^{10}L_{\sun}$
to a fixed mass, we used the  $M/L_{\rm bol}$ tabulated by the EPS models.
For $[{\rm M/H}]=-1.27$, a typical value for the globular clusters under consideration, $N_{\rm eff}$ values for a $10^6$~M$_{\sun}$ 
population are 300 (15 Gyr),  270 (10 Gyr),  and 260 (8 Gyr). These suggest that clusters
with masses of $10^6$~M$_{\sun}$ should not have their integrated mid-IR photometry
strongly affected by stochastic effects, while clusters with mass  $10^5$~M$_{\sun}$ would
have $N_{\rm eff}$ well below the requirements.
We used this result to set our lower mass limit at $10^6$~M$_{\sun}$.

\subsection{Cluster Sample Selection}

For this first comparison between SSP models and star cluster colors, we wanted 
to remove age as a variable if possible, and select only old clusters.
The clusters we considered all belong to the Milky Way, M31, or the LMC.
While many other Local Group dwarf galaxies have a few globular clusters each
\citep[see the compilation by][]{forbes00}, the few available estimates of those clusters' masses in the literature
show the largest of them to have masses about $10^5$~M$_{\sun}$.
M33 has numerous star clusters, with a recent catalogue given by \citet{sarajedini07}; however it has few truly old clusters,
and few clusters with mass estimates comparable to those available in other galaxies.
For Milky Way globular clusters we take the inclusion of a cluster in the catalog of \citet{harris10} 
to mean that its age is $\gtrsim10$~Gyr. For LMC clusters
we considered only clusters classified as  having ages $>10$~Gyr by 
\citet{pessev08}; these ages were compiled from the results of color-magnitude diagram (CMD) fitting.
Age determination for M31 clusters is more difficult, and, with 
one exception \citep{brown04}, CMD-fitting ages from the main sequence turnoff
are not available.
\citet{perina11} noted that deriving M31 cluster ages from integrated spectroscopy can be problematic.  
They found that two clusters previously suspected to be of intermediate age based on integrated spectra
had blue horizontal branches and were likely of comparable age to the oldest Galactic GCs.
They also found significant differences between ages determined from spectral energy distribution (SED) fitting
and the CMD analysis. With the caveat that ages from spectroscopy may be unreliable,
a uniform determination of ages for the M31 GCs reduces scatter due to 
mixing of different methods, and we use the values by \citet{caldwell11} as the basis for selecting our sample. These
were derived by comparing Lick indices from integrated spectroscopy to the SSP models of \citet{schiavon07}. An  important
limitation of this analysis is that ages could not be determined for clusters with
metallicities ${\rm [Fe/H]}\lesssim -1$. It is possible that some younger clusters
could be lurking in the sample, although we considered only M31 clusters that
had not been identified as young by any previous work.

There are a number of methods for estimating the total mass of a star cluster.
While velocity dispersions either from integrated spectra or 
spectra of individual stars provide the most direct measurements of the
gravitational potential of a cluster, such measurements are not available for
many Local Group star clusters, particularly in M31 which has the largest population.
A more widely-applicable method involves measuring the total luminosity and multiplying
by a mass-to-light ratio. Total luminosity  can be measured by either large-aperture photometry
or fitting and extrapolation of a surface brightness profile. 
For M31 clusters, a comprehensive set of mass estimates is available in the work of 
\citet{caldwell11}, who estimated masses by multiplying reddening-corrected $V$-band
luminosities \citep[from][]{caldwell09} by $M/L_V=2$. 
With the goal of selecting a uniform sample, we used these mass estimates exclusively to select the final M31 sample;
a few additional outlying clusters for which masses were not given by \citet{caldwell11} but which have
mass estimates in their online catalog (G001, B358, MGC-1, B468, B293) were also included.
For Milky Way and LMC clusters, we used  near-infrared $K_s$-band magnitudes
as given by \citet{cohen06} and \citet{pessev06, pessev08}.
The near-infrared is often used as an indicator of stellar mass in studies of
external galaxies since it is expected to be dominated by low-mass stars and is not strongly sensitive to
recent star formation \citep{bell01}.
The clusters' magnitude were converted to luminosities using $M_{\sun, Ks}=3.29$ \citep{blanton07} 
and multiplied by $(M/L)_{K_s}=0.9$ in solar units, the lower end of the range quoted for
single-metallicity populations by \citet{maraston05}. We also
considered cluster masses derived from fitting of optical surface brightness 
profiles  by \citet{mclaughlin05}; these masses are
primarily derived from $V$-band observations (corrected for extinction) and assume
$(M/L)_V=2$ \citep{mclaughlin00}. Some Milky Way clusters (particularly core-collapse
candidates) are missing from the \citet{mclaughlin05} database.

We combined the two methods to select a final list of massive 
Local Group globular clusters.  Excluding the anomalous cluster $\omega$ Cen, only 
a handful of Milky Way clusters have masses $>10^6$~M$_{\sun}$: \object{NGC~104}, 
\object{NGC~6388}, \object{NGC~6441}, and  \object{NGC~6715}. A further 8 Milky Way
clusters (\object{NGC~2808}, 
 \object{NGC~7089},
 \object{NGC~2419},
 \object{NGC~6266},
 \object{NGC~6273}, 
 \object{NGC~6356}, 
 \object{NGC~6402},
 \object{NGC~6440})
and two Magellanic Cloud clusters (\object{NGC~1835}, \object{NGC~1916}) 
had masses likely to be above $5\times10^5$~M$_{\sun}$.
But nearly all of the massive Local Group globular clusters belong to M31, a total of 58 objects.
After examining the available {\it Spitzer}/IRAC data for Milky Way and LMC clusters, we 
eventually decided against using data on those galaxies'  GCs  in our analysis: the Milky Way cluster 
data had technical issues (long frametimes and in-place repeats) as well as 
only covering  a 
single $5\arcmin \times5\arcmin$ field of view in all four IRAC bands.
The LMC clusters NGC~1916 is located near a region of diffuse emission, making background subtraction
for this cluster extremely difficult.
IRAC photometry of a larger sample of LMC clusters has been reported  by \citet{pessev10}.

\subsection{IRAC Observations and Photometry}

The M31 bulge and some of the brightest M31 clusters had individual deep   {\em Spitzer}/IRAC observations as
part of PID 3400 (PI M.\ Rich). These images also include a number of additional clusters
fortuitously located near the target clusters. The data were taken with 30-second
frame times, 22 dither positions with 4 frames per position, and so
have overall depths of about 40 minutes per sky position.
Saturation limits for IRAC images with these frametimes are in the range $8<m_{\rm Vega} < 11$
\citep{iracihb}.
The brightest GCs in M31 have $K_s \sim 11$ \citep{rbc04}, 
so saturation of the clusters was not expected to be an issue.
We retrieved the data from the  {\em Spitzer} Heritage Archive and used the
post-BCD mosaics as produced by versions 18.7 and 18.18 of the pipeline.

\clearpage
\begin{figure}
\includegraphics[width=15cm]{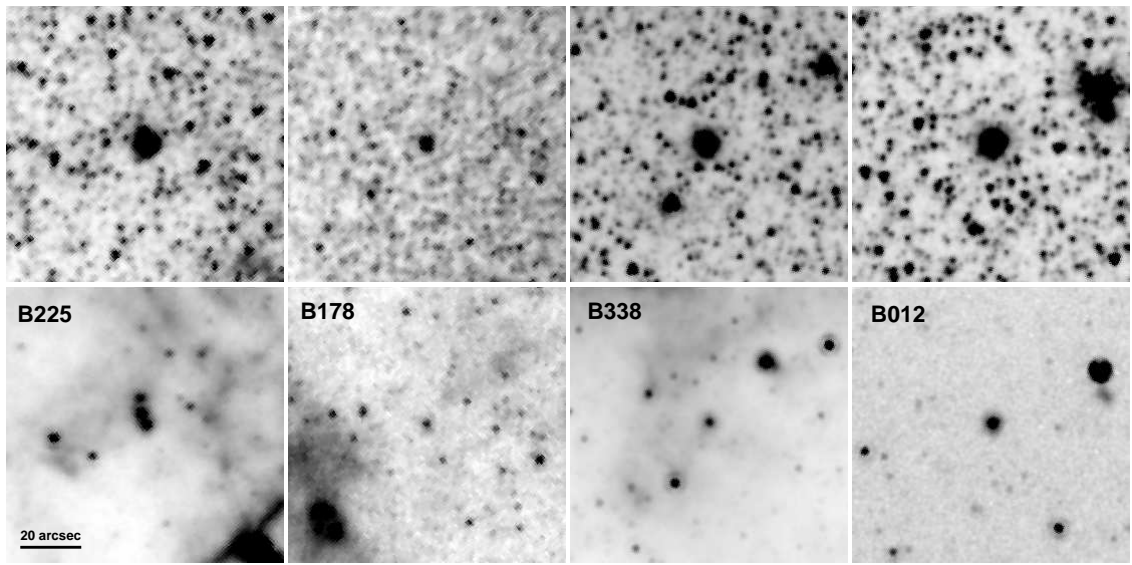}
\caption{IRAC 3.6\micron{} (top) and 8.0\micron{} (bottom) images of M31 globular clusters.
All images are 90\arcsec{} square with north up and east to the left.
The clusters range from $K_s\sim 11.5$ (B225) to $K_s\sim13.3$ (B178).
The left two panels show clusters in the shallow M31 mosaic (PID 3126) while the right two
panels show clusters observed in deeper images (PID 3400). 
\label{fig:m31_img}}
\end{figure}

\clearpage

The main observational dataset for the M31 clusters
comprises the IRAC mosaics taken as part of program PID 3126,
described in \citet{barmby06a}. In brief, the 
observations cover a total area of about $3\fdg7 \times 1\fdg0$
along the major axis of M31, with an extension to include NGC~205. 
The central portion of the disk was covered  
to a depth of six 12-second frames per position, and the 
outer area by four 30-second frames per position.
The mosaics used for this analysis 
are improved versions of the mosaics used by \citet{barmby06a}
and are described by \citet{mould08}.
Of the 58 clusters in our list of massive M31 globulars, 11 were imaged
in all four bands in the deep PID3400 data, 41 were imaged only on the shallow
M31 mosaics, two objects (B405, MGC-1) had no IRAC images available, 
and 3 objects (B240, B381, B383) were at the edge of the mosaic image and had only
noisy data in 4.5 and 8.0~\micron\ available.
Cluster B095 is located about 7\arcsec\ from a much brighter star and is contaminated by
detector artifacts due to the star; we judged that obtaining
reliable photometry in the IRAC bands was not possible.
The final sample of M31 clusters analyzed here therefore has a total of 52 members.
Sample IRAC images of M31 globular clusters are shown in Figure~\ref{fig:m31_img}.

\begin{figure}
\includegraphics[width=15cm]{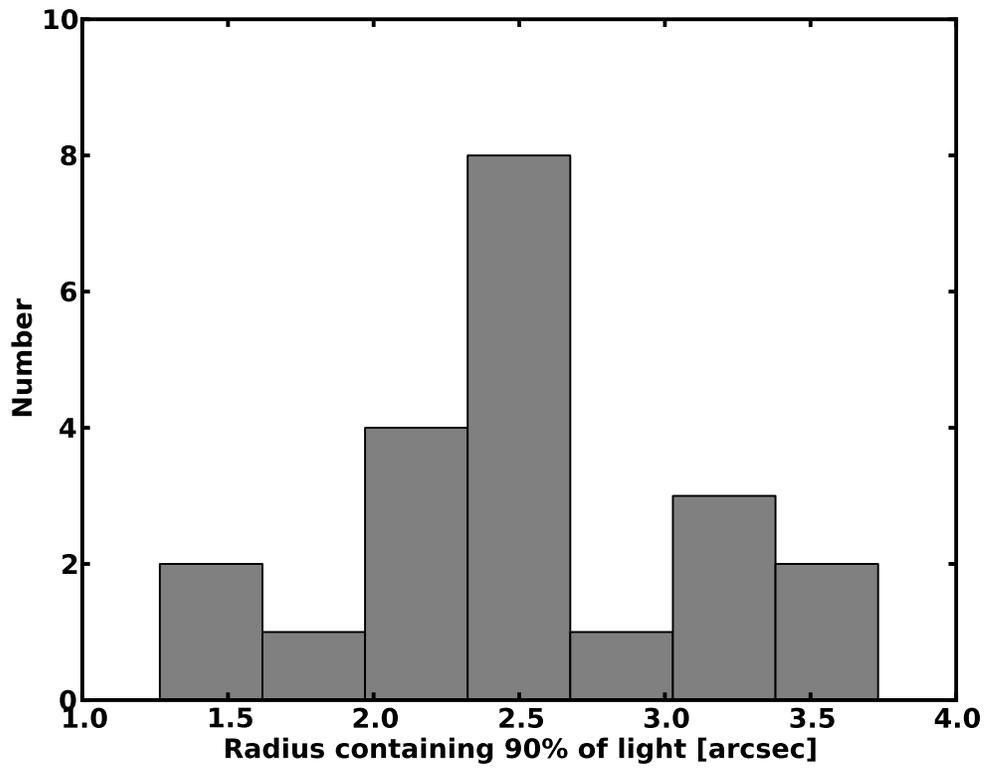}
\caption{Distribution of $R_{90}$---the radius containing 90\% of the integrated light,
in projection--- for M31 globular clusters.
\label{fig:r90}}
\end{figure}

The typical half-light radius of an M31 GC is $R_h\sim 3$~pc, or about
0.75\arcsec\ \citep{peacock10a, barmby07}. 
While this is smaller than the full-width at half-maximum of the IRAC point spread function (PSF), 
it is large enough that M31 GCs appear slightly resolved (see Figure~\ref{fig:m31_img}).
Aperture photometry is thus more appropriate for these objects than PSF-fitting 
photometry. But how do we best account for the different PSF in the
different IRAC bands, and decide which aperture should be used?
From the point of view of capturing all of the light from a cluster, 
a large aperture is desirable. Use of `total' magnitudes also facilitates
comparison between widely-separated wavelength regimes, where not
only the PSF but the apparent cluster structure, may differ
\citep[see][for a discussion of the subtleties in aperture-matching]{cohen06}.
However, the use of a large aperture would result in low signal-to-noise
in 5.8 and 8.0~$\mu$m\  bands, where the clusters are faint  and the background
bright. Using smaller apertures has the problem that the different PSF results in
aperture corrections that are not the same in all bands.

To deal with the aperture issues, we took two approaches to photometry of the M31 clusters.
To compute IRAC-only colors, we computed aperture magnitudes in a relatively small
apertures, and dealt with the variable PSF by convolving the images
in the 3.6, 4.5, and 5.8~$\mu$m\  bands to the 8.0~$\mu$m\  resolution. \citep[A similar approach
was used by][in their study of near-to-mid-infrared colors of Coma cluster galaxies.]{clemens09} 
For colors which combine the 3.6 and 4.5~$\mu$m\  bands with visible-light or near-infrared
bands, we measured `total' IRAC magnitudes in a standard large aperture.
Convolution was done using the IRAF {\sc fconvolve} task and the convolution kernels
described by \citet{gordon08}.  For the clusters observed as part of the large mosaics, we extracted $2.5\arcmin \times 2.5\arcmin$ 
`postage stamp' images and convolved these, rather than attempting to convolve the
original mosaics. We compared  aperture photometry on the 
original and convolved deep images to show that the convolution performed as
expected: the median difference in aperture magnitudes between convolved and unconvolved
images was close to the difference between aperture corrections between the 
convolved band and the 8.0~$\mu$m\  band  as tabulated in the IRAC Instrument Handbook \citep{iracihb}.
To choose a small aperture size, we used  the results of surface brightness profile fitting
from {\it Hubble Space Telescope} images of M31 globular clusters \citep{barmby07} to
compute $R_{90}$, the projected radius containing  90\% of the integrated light, for each cluster. 
Figure~\ref{fig:r90} shows the distribution of $R_{90}$ for the
34 clusters from our sample analyzed by \citet{barmby07}.
The median of the $R_{90}$ distribution is 2\farcs5, with a maximum of 3\farcs7. 
These values refer to the intrinsic profile before convolution with the \textit{HST} PSF; the 
effects of the  \textit{Spitzer/IRAC} PSF are more significant and mean that $R_{90}$ as observed
in IRAC images will be larger. To maximize signal-to-noise and still include a significant fraction
of cluster light, we chose one of the standard IRAC aperture sizes, a radius of 3\farcs66.

Aperture photometry was performed using the IRAF/APPHOT package.
Flux was measured in circular apertures of radii 3\farcs66 (and 12\farcs2 for the 3.6 and 4.5~$\mu$m\  bands only),
with background measured in an annulus of inner and outer radii 3\farcs6--8\farcs5  (14\farcs6--24\farcs4)
subtracted. 
This background  annulus for the small aperture 
is the smaller of two tabulated in the Instrument Handbook. 
We used it because a few clusters had nearby diffuse emission which could bias the 
background estimate in a larger annulus.  We verified that in most cases
the choice of  background annulus had no effect on the resulting photometry.
Calibration to the Vega system was made using the Vega IRAC-band flux
densities given by the Spitzer Science Center.
Uncertainties on the photometric measurements were estimated using
the formula in {\sc apphot/phot}, with a correction to the background
noise term which accounts for the correlated noise introduced by pixel
re-sampling during mosaicing. As the background variation is by far the dominant
source of noise in these measurements, this is an important correction.
Table~\ref{tab:gcphot} gives the photometry.
We checked the internal consistency of the photometry
by comparing measurements of the same clusters on
the deep and shallow IRAC images. 
There is some scatter between the two
image types, but magnitude offsets are consistent with zero.
In the 4 IRAC bands, the offsets are (mean $\pm $ standard deviation): 
$-0.03\pm0.04,-0.06\pm0.08,-0.06\pm0.07,-0.06\pm0.09$.

\subsection{2MASS Photometry and Extinction Correction}
\label{2massphot}

While globular clusters have not been extensively observed in the mid-infrared, there is
a long history of using them to test population synthesis models in the near-infrared,
so it is of interest to connect these two regimes.
The most widely-used near-infrared photometric
system is that of 2MASS \citep{skrutskie06}, and the 2MASS data on M31 GCs has been examined by
\citet{rbc04} and \citet{nantais06}. \citet{peacock10a} also presented observations of M31 GCs in 
the $K_s$ band based on deeper UKIRT imaging. We chose to re-measure the M31 GC
magnitudes on the 2MASS `6x' Atlas images  because the observations of \citet{peacock10a}
only covered 60\% of our cluster sample,  the compilation of \citet{rbc04} used the original
shallow 2MASS images of M31, and the compilation of \citet{nantais06} used the 2MASS
Point Source Catalog,  which might be inappropriate for the larger M31 clusters.
To maintain consistency
with the colors measured using the IRAC data, we retrieved 2MASS  images of the M31 GCs
and used these to measure aperture magnitudes with the same large photometric and background apertures
used for the IRAC data. We chose to use the large apertures rather than trying to match the
spatial resolution between IRAC and 2MASS because  the temporal variation
in the 2MASS PSF makes any such matching significantly more complicated.
The photometric uncertainties were estimated using the same
procedure as for the IRAC magnitudes, which is based on the procedure detailed
in section 8av of the 2MASS All-Sky Data Release Explanatory Supplement.
Table~\ref{tab:gcphot} gives the 2MASS $K_s$ photometry.
We compared our aperture photometry with the
total magnitudes measured on UKIRT images by \citet{peacock10a}. The UKIRT 
magnitudes are slightly fainter, with the median offset $0.08\pm0.09$~mag.
This is not too surprising, since we used an aperture slightly larger than most
of the apertures used by  \citet{peacock10a}. It could result in
our `total' $K_s$ magnitudes being slightly too bright; however for the purposes of
computing 2MASS-to-IRAC colors this effect should cancel out.

While extinction has a much smaller effect on photometry in the mid-infrared  than in
visible-light or even the near-infrared, the effect is not completely negligible.
The exact shape of the extinction law in the mid-infrared appears to vary with
position in the Galaxy
\citep{indebetouw05,roman-zuniga07,flaherty07,nishiyama09,gao09} but
all of these analyses agree that the extinction at IRAC wavelengths is greater than
would be expected from a power-law extrapolation of the near-infrared extinction curve.
In the analysis that follows, we correct the cluster colors for extinction
using the $E(B-V)$ values given by \citet{caldwell11}.  These values include both
foreground extinction and extinction internal to M31. 
For this analysis we use the bandpass-averaged extinction values of
\citet{indebetouw05}, as tabulated in \citet{gao09}: $A_{3.6}/A_{K_s} =0.56$
and $A_{\lambda}/A_{K_s} =0.43$ for the other three IRAC bands. We assume $A_{K_s}/E(B-V)=0.35$
\citep{cambresy02} and the standard $A_V/E(B-V)=3.1$.

\subsection{Population Synthesis Models}

While there are numerous population synthesis models currently available ---
two of the most widely-used being \citet{bruzual03} and \citet{maraston05} --- to our knowledge only 
three models give explicit predictions for magnitudes in the IRAC bandpasses.
These are GALEV \citep{kotulla09}, FSPS \citep{conroy09, conroy10}, and the SSP models
provided by the Padova group through their STEL webserver\footnote{\url{http://stev.oapd.inaf.it/}} \citep[hereafter referred to as `Padova']{marigo08,girardi10}.
While many other models can be used to generate predicted IRAC magnitudes from
spectral energy distributions, using three model families suffices for this first analysis.
These models contain some of the same input ingredients
but differ in other ways, such as the treatment of thermally-pulsing asymptotic giant branch and
horizontal branch stars.  GALEV also provides chemically consistent modeling of galaxies 
while the other two do not, although this difference is not relevant for the present study.

Population synthesis models have mostly focused on the visible-light regime, so
it is reasonable to ask what the input ingredients of these models are in the infrared.
The treatment of red giant and asymptotic giant branch stars is expected to be particularly important, since 
these evolutionary stages dominate the light of old populations, especially in the near-infrared
\citep[][and many others]{buzzoni89, maraston05}.
All three models were consider here use Padova isochrones, so an important factor in any different
outputs  is the stellar libraries. GALEV and FSPS both use the 
BaSeL stellar libraries: version 2.2 for GALEV \citep{lejeune98} and version 3.1 \citep{westera02}
for FSPS. BaSeL is based mainly on the stellar atmosphere models of
\citet{kurucz92}, but those models do not include the coolest stars, so BaSeL
also includes model M giant atmospheres from \citet{fluks94} and \citet{bessell89, bessell91}.
The Padova SSP models use some of the same input ingredients---newer ATLAS models
from \citet{castelli03} and the  \citet{fluks94} models---but without going through the BaSeL
library. Since the \citet{fluks94} models extend to wavelengths up to 12.5\micron{} and the 
\citet{bessell89, bessell91} models only to 4.1\micron, we can infer that the former models
provide the basis for the predictions tested here; however, this is not explicitly stated in
any of the model descriptions. 
The thermally-pulsing asymptotic giant branch can also be an important contributor to the integrated light
of star clusters \citep{maraston05}, although it is not expected to be important at globular 
cluster ages.
For these stars, FSPS uses empirical spectra from \citet{lancon02} while the
Padova models use synthetic spectra from \citet{loidl01}.  
Both the \citet{lancon02} and  \citet{loidl01} spectra extend only to 2.5 \micron.  
The Padova models extend the AGB spectra beyond 2.5~$\mu$m\  with a Rayleigh-Jeans model:
FSPS extrapolates carbon star spectra with \citet{aringer09} synthetic spectra.
\citet{kotulla09} did not describe any special treatment of TP-AGB stars in GALEV.

For all three models, we generated predicted SSP colors in the 2MASS and IRAC bands for a 
\citet{salpeter55} 
($0.1-100$~M$_{\sun}$) initial mass function%
\footnote{Comparison of all three models' predictions for  \citet{salpeter55} and \citet{kroupa01} 
initial mass function showed that this choice of IMF has a negligible effect on predicted colors of old
stellar populations.} and ages from 1 Gyr up to the oldest available.
We retrieved SSP predictions from the online
GALEV interface using all available fixed metallicity values  ${\rm [M/H]}= \log(Z/Z_{\sun}) = (-1.7, -0.7, -0.3, 0.0, +0.3)$.
For FSPS, we used version 2.3 of the code to generate predicted magnitudes
for simple stellar populations with metallicity values ${\rm [M/H]}=(-1.68,  -0.69, -0.30, 0.0, +0.20)$.  
The default options for  other parameters were used.
For the Padova models, we used metallicities of 
${\rm [M/H]}=(-1.68,-0.70,-0.30, 0.0, +0.20)$, with the default parameters: evolutionary tracks
from \citet{marigo08}, with the \citet{girardi10} case A correction, and no circumstellar dust (although see below).
For all models, no intrinsic dust extinction was applied. 
We interpolated the GALEV magnitudes to ${\rm [M/H]}=+0.2$ so that all three models
represented the same five metallicity values.  
These metallicity values adequately cover 
the range of metallicities for clusters in the present sample ($-1.9\leq{\rm [M/H]}\leq -0.1$); 
although M31 has lower-metallicity clusters \citep{caldwell11}, they are not present among the most massive objects.

\clearpage
\begin{figure}
\includegraphics[width=15cm]{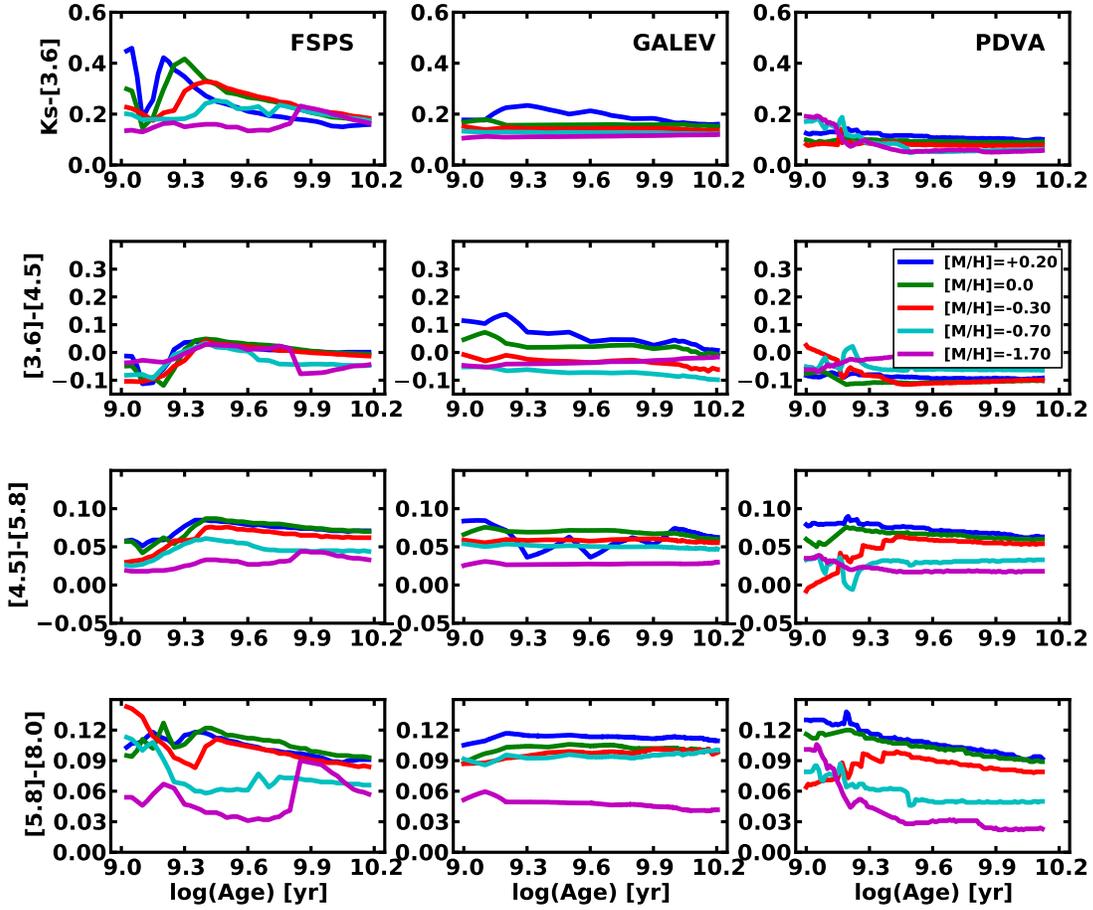}
\caption{Predicted IRAC colors from three population synthesis models,
as a function of simple stellar population age. Left column: FSPS \citet{conroy09},
center column: GALEV \citep{kotulla09}, right column: Padova \citep{marigo08}.
The same set of metallicities is plotted for all 3 models.
\label{fig:mod_age}}
\end{figure}
\clearpage

Figure~\ref{fig:mod_age}  shows predicted colors from the models as functions of age.
At old ages,  all models show a rather small range of predicted  mid-infrared colors, only a
few tenths of a magnitude.  Although the colors mostly become slightly bluer with age, 
the size of this effect is very small.
There are some systematic differences in the models' predictions, with, in general, the 
FSPS models predicting the greatest color variations over the displayed age range.
While there is general agreement that higher metallicities result in redder colors within
a model family, the different models do not agree on either the size of this effect or its location.
For example, the FSPS $K_s-[3.6]$ color for the most  metal-poor model
is redder than the same color for the most metal-rich Padova model. 
Each of the models has at least one metallicity track that behaves differently from 
others in the same family, but which metallicity is the outlier changes depending 
on the color and model family. Since all three models
use the same set of isochrones, the difference in predicted colors must be due to some other
factor. One likely contributor is the different spectral libraries used for post-main-sequence stars.
Another could be numerical issues surrounding  the stability of integration over
short-lived, but luminous stages \citep{maraston05, tinsley76}.

\begin{figure}
\includegraphics[width=15cm]{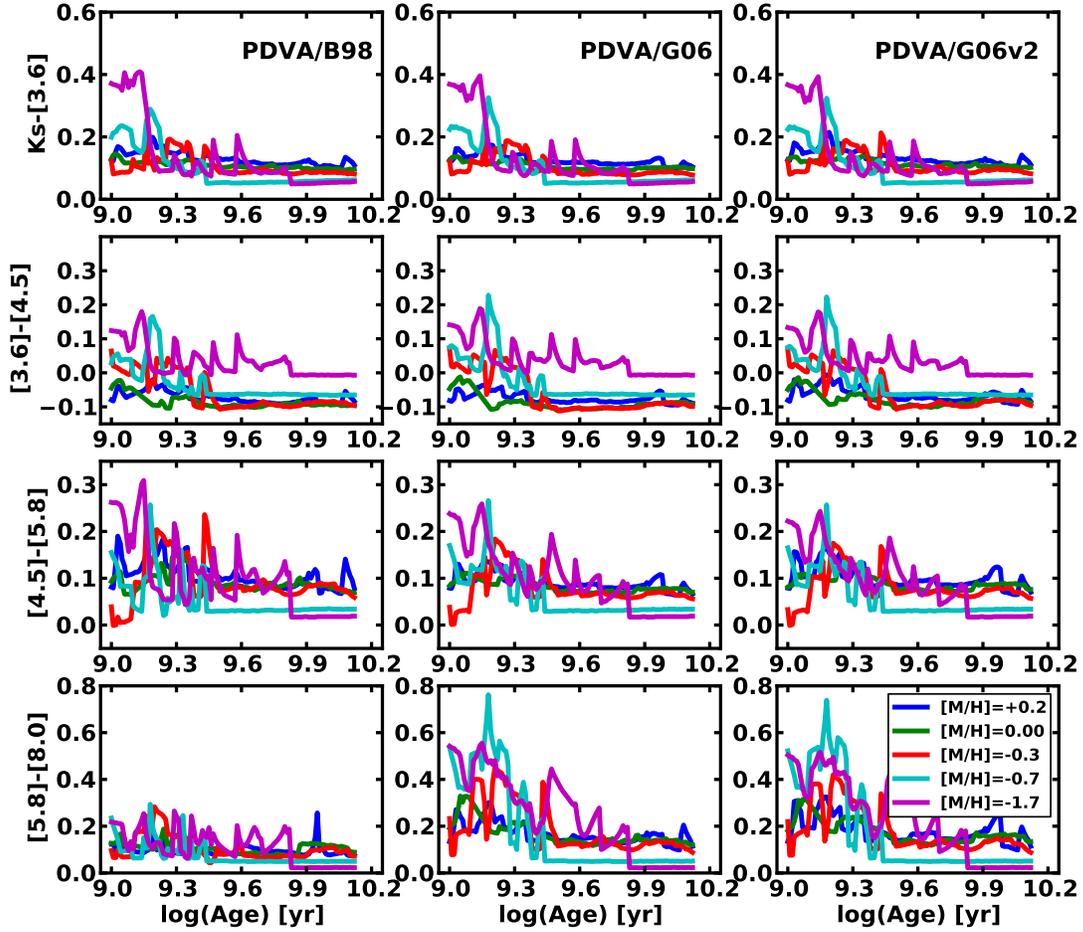}
\caption{
Predicted IRAC colors from Padova population synthesis models,
as a function of simple stellar population age. The 3 columns show
colors resulting from 3 different options for circumstellar dust:
B98 is \citet{bressan98}, G06 is \citet{groenewegen06} 100\% AlO$_{\rm x}$
for M stars and 100\% amorphous carbon for C stars, and G06v2
is \citet{groenewegen06} 60\% silicate/40\% AlO$_{\rm x}$
for M stars and 85\% amorphous carbon/15\% SiC for C stars.
\label{fig:dustymod_age}}
\end{figure}

While Milky Way globular clusters have not shown evidence for intracluster dust
\citep{matsunaga08,barmby09}, mid-infrared observations have provided important
evidence for dust in the envelopes of individual mass-losing stars
 \citep{boyer10,origlia10}. So it is reasonable to ask what the effects of 
 dusty mass loss would be on the integrated mid-infrared colors of a clusters.
The Padova models can optionally include these effects for stars with significant mass loss:
red supergiants, TP-AGB, and upper-RGB stars \citep[see][for details]{marigo08}.
Figure~\ref{fig:dustymod_age} shows the result of adding circumstellar dust to the Padova
simple stellar populations. 
This figure shows the Padova models with three different prescriptions for circumstellar dust: 
silicate dust for M stars and carbon-rich dust for  C stars \citep{bressan98}, 
100\% AlO$_{\rm x}$ for M stars and 100\% amorphous carbon for C stars,  and 
60\% silicate/40\% AlO$_{\rm x}$ for M stars and 85\% amorphous carbon/15\% SiC for C stars 
\citep[the latter two both from][]{groenewegen06}. 
Comparing with the right panel of Figure~\ref{fig:mod_age} shows that
including circumstellar dust emission causes the colors to vary more with age
(all of the Padova models are plotted with the same time resolution) and become 
redder, particularly at the longer wavelengths. These effects are stronger for ages 
$\lesssim 8$~Gyr: models at the oldest ages are close in color to the non-circumstellar-dust
version. In most cases, the choice of circumstellar dust prescription does not have a large effect
on the predicted colors. The exception is $[5.8]-[8.0]$ where the \citet{bressan98}
prescription results in bluer colors, particularly at lower metallicity,  than either
of the two prescriptions from \citet{groenewegen06}. 
To maintain compatibility with the other two families, the dusty
circumstellar models are not shown in the comparison with the cluster data,
but these effects will be considered in the discussion.

\section{Analysis and Discussion}

\subsection{Cluster color distributions} 

\begin{figure}
\includegraphics[width=15cm]{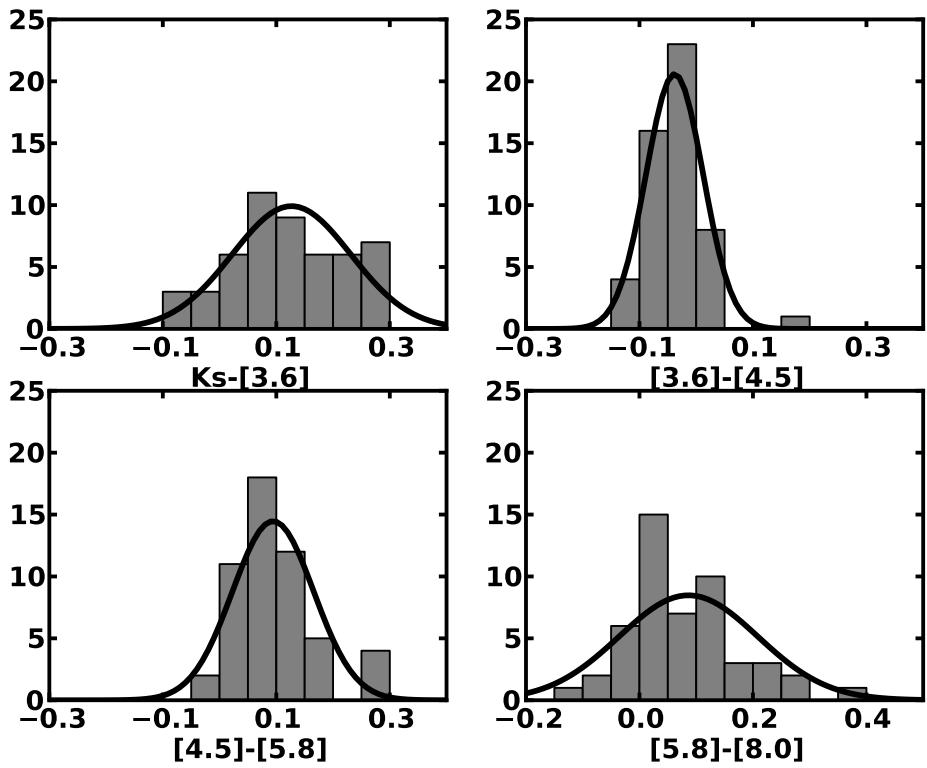}
\caption{Distribution of near- and mid-infrared extinction-corrected colors for massive M31 globular clusters.
Gaussians with mean and standard deviation computed from the data are overplotted
for reference.
\label{fig:col_dist}}
\end{figure}

Before comparing to models, the most basic analysis of the globular cluster colors
is to examine their distribution. Figure~\ref{fig:col_dist} shows the distribution of
the four colors constructed with bandpasses adjacent in wavelength. Overplotted are
Gaussians normalized to the same area, with the data-derived means and standard deviations:
$K_s-[3.6]=0.13\pm0.11$, $[3.6]-[4.5]=-0.04\pm0.05$, $[4.5]-[5.8]=0.09\pm0.07$ 
and $[5.8]-[8.0]=0.09\pm 0.12$ (mean $\pm$ standard deviation).
\citet{pahre04} gave  theoretical colors of M0 III stars in the IRAC bands of
$[3.6]-[4.5]=-0.15$, $[4.5]-[5.8] = +0.11$ and $[5.8]-[8.0]=+0.04$. 
The mean colors of the cluster sample are  formally consistent with the naive expectation of zero magnitudes,
and in the longer wavelengths  with the theoretical colors. The mean $[3.6]-[4.5]$ color of the clusters 
is slightly redder  than the expected giant color, as are the model colors described in
Section~\ref{modcol}, so we suggest that
there may be a problem with the value given by \citet{pahre04}. 
The $[3.6]-[4.5]$ color shows the narrowest color distribution, unsurprising
since these two bands have the most similar PSF and do not suffer from
any possible aperture-matching problems between 2MASS and IRAC data.
At the longer wavelengths, where the clusters
are fainter and the background brighter, the observational uncertainties are larger.
Contributions by diffuse PAH emission could also
skew the 5.8 and 8.0~\micron{} fluxes, although this should be minimized by the 
close-in background aperture used for photometry.

As Figure~\ref{fig:mod_age}  shows, the expected range of mid-infrared
colors for old stellar populations is rather small, a few tenths of a magnitude. 
Are the distributions that we measure the result of a single intrinsic color, broadened
by observational uncertainty?  One argument against this is that, for the IRAC-only colors, the standard deviations of the 
color distributions are broader than the average photometric uncertainties, by factors of 1.5--2.5. 
Applying the D'Agostino-Pearson normality test to all four
color distributions, we can reject the hypothesis of normality in the $[3.6]-[4.5]$ and $[5.8]-[8.0]$ colors,
but not in $K_s-[3.6]$ or $[4.5]-[5.8]$. We also compared the color distributions to Gaussians with
zero mean and standard deviation given by the average photometric uncertainty using 
a Kolmorgorov-Smirnov test. In all cases the hypothesis that the observed distributions were drawn
from the zero-mean distribution was rejected. We conclude that the cluster colors have some intrinsic
spread and that the observed distributions are not merely delta functions broadened by observational uncertainty.
Mis-estimates of reddening should have only a small effect on colors derived from adjacent band and
are unlikely to cause significant scatter. A spread in ages is a possible reason for a color distribution spread; 
this is discussed further in Section~\ref{modcol}. Another possible explanation is that the relatively poor
spatial resolution of IRAC leads to contamination of some clusters' colors by sources nearby in the M31 disk.

\clearpage

\begin{figure}
\includegraphics[width=15cm]{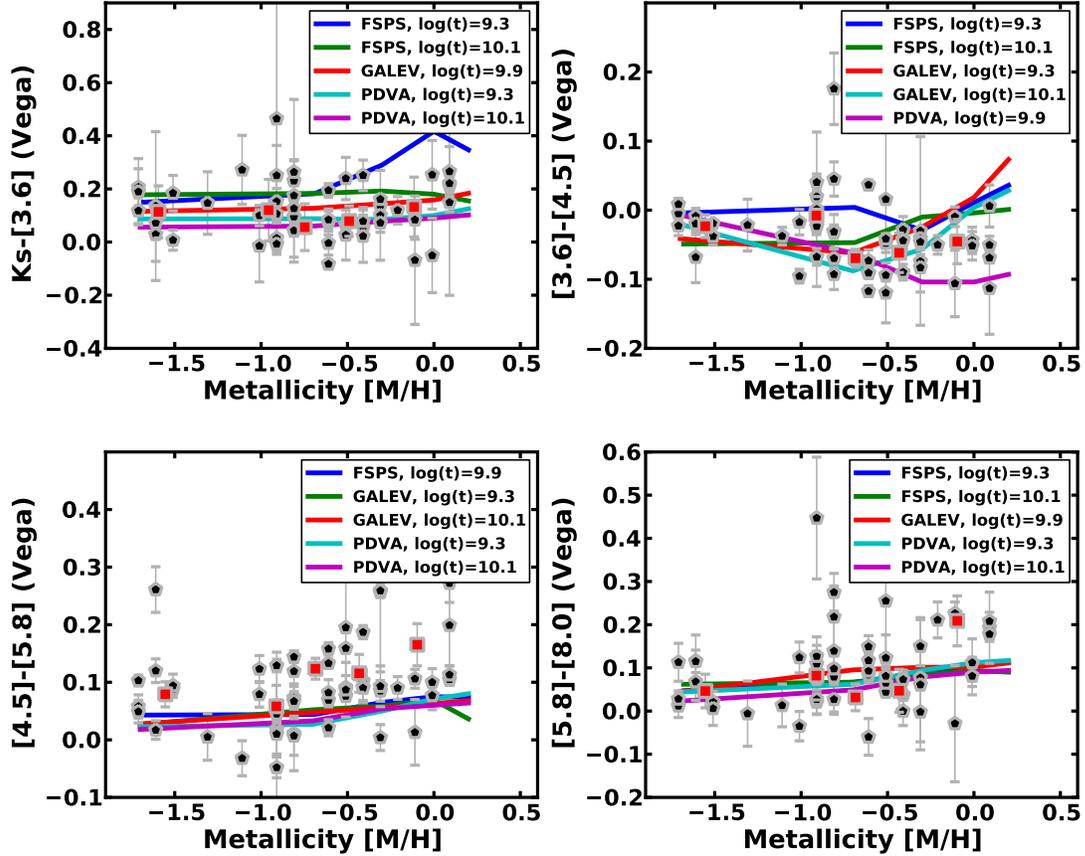}
\caption{Near and mid-infrared colors (corrected for extinction) as a function of metallicity
for massive M31 globular clusters (points) and SSP model predictions (lines). Each of the
model prediction lines represents one model family at a specific age between 2 and 12.5 Gyr.
Solid black symbols
are individual M31 clusters; open symbols (red in the online version) represent the luminosity-weighted mean colors
of clusters in quintiles of metallicity.
\label{fig:col_met}}
\end{figure}
\clearpage

\subsection{Observed and model colors}
\label{modcol}

While, by design, the M31 clusters in our sample cover a fairly small range in mass,
they do cover a range of about 2 dex in metallicity. Comparing models to cluster colors as a function of
metallicity allows us to see how the ranges of the data and models coincide.
Figure~\ref{fig:col_met} shows these comparisons for the same colors  as shown in Figure~\ref{fig:col_dist}. 
The models plotted are a subset of those shown in Figure~\ref{fig:mod_age}, chosen to
represent the range of model colors for ages of 2, 8, and 12.5 Gyr.
For metallicity values of the clusters, we used the spectroscopic ${\rm [Fe/H]}$ measurements of \citet{caldwell11}
and converted to total metallicity with ${\rm [M/H]}={\rm[Fe/H]} + 0.94[\alpha/{\rm Fe}]$ \citep{thomas03}.
\citet{beasley05} found a large spread in the degree of $\alpha$-element enhancement for M31
clusters; we use a typical value of $[\alpha/{\rm Fe}]=0.2$.
As well as data for individual clusters, the Figure also shows the luminosity-weighted mean colors
of clusters grouped in quintiles of metallicity. In general, the clusters cover a wider
range of color than do the models, but the models typically fall within the range of the cluster data.
The reddest model in the $K_s-[3.6]$ and $[3.6]-[4.5]$ plots is the FSPS 2~Gyr model,
which does not appear to be a good fit to the data; however, the 2 Gyr Padova model
does match the data in $K_s-[3.6]$, as does the 2 Gyr GALEV model in  $[3.6]-[4.5]$.
The observational scatter in these two colors is large enough that older models are also 
broadly consistent with the data. 

At the longer IRAC wavelengths,  the range of model colors is very small
and the models are generally in good agreement with each other.
In $[4.5]-[5.8]$, the cluster mean colors are slightly redder than
all of the models by up to 0.1~mag, but the offset is  statistically significant
in only two of the five metallicity bins. The scatter of some objects to redder
colors could be due to systematic effects in the photometry (incomplete
background subtraction, failure to completely match apertures with the convolution
method), or, more interestingly, it could indicate that some clusters' colors are affected by 
circumstellar dust emission. 
In $[5.8]-[8.0]$, the mean cluster colors are generally in agreement with the models except possibly 
at the highest metallicity.
For the most part, the redder colors predicted by the Padova models for stars with circumstellar dust
do not appear to match the cluster data. This lends support to the idea that these clusters
are truly old objects and their infrared light is not dominated by emission from circumstellar dust.
 
Optical to near-infrared colors (e.g., $V-K_s$) are good metallicity indicators in old stellar
populations, because they are sensitive to the temperature of the red giant branch.
Optical-to-mid-infrared colors have been less well studied.
\citet{spitler08} measured $V-[3.6]$ for GCs in NGC~5128 and NGC~4594
and  found that this color showed  a  good correlation with spectroscopic metallicity.
To see if the same is true for M31 GCs, we computed optical-to-mid-infrared 
colors using the $V$-band photometry tabulated by \citet{caldwell09},
corrected for extinction using the $E(B-V)$ values given by \citet{caldwell11}. 
These colors are much more sensitive to reddening than the infrared-only colors
considered above.  \citet{caldwell11} estimated the typical reddening uncertainty to be 0.1~mag:
changing a reddening estimate by this much would correspond to a change of nearly 0.3~mag 
in  reddening-corrected  $V-K_s$. 

\clearpage

\begin{figure}
\includegraphics[width=15cm]{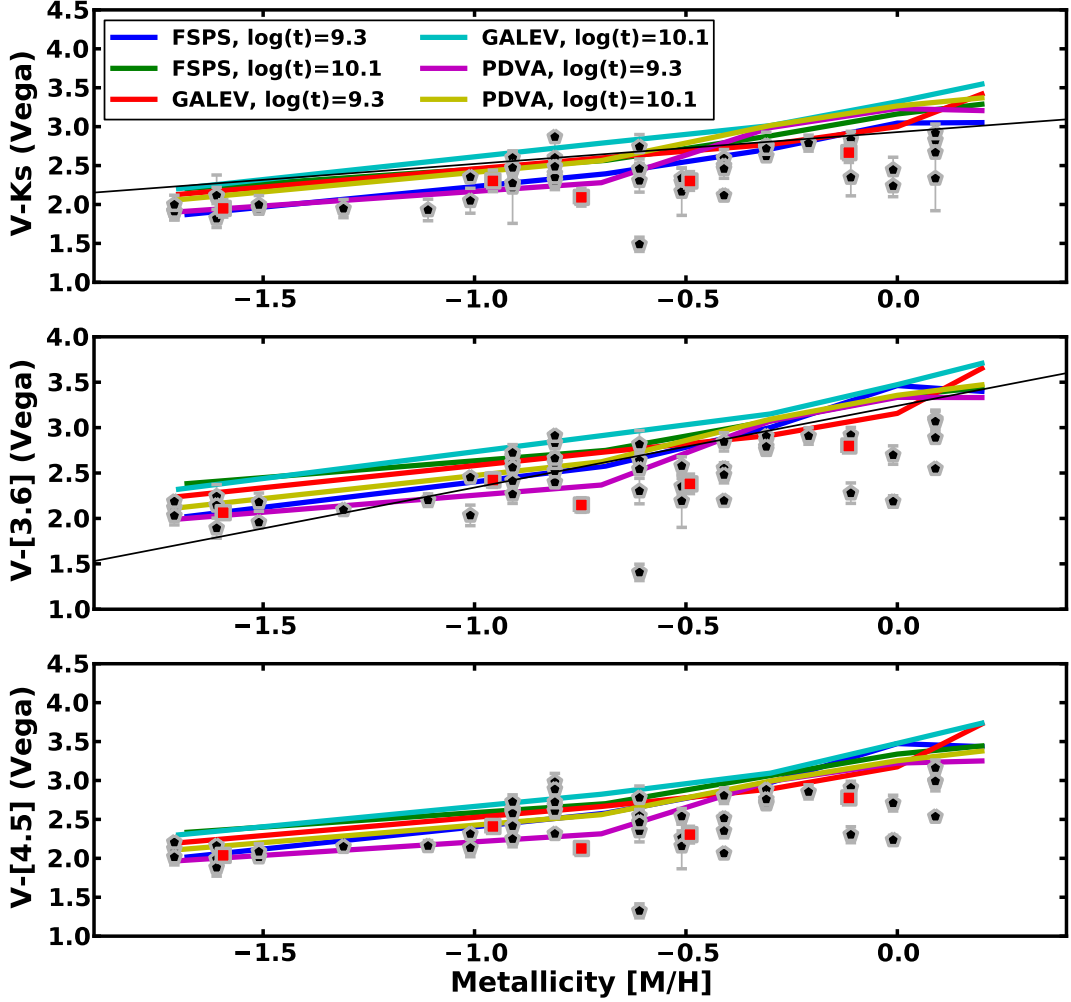}
\caption{Optical to near/mid-infrared colors (corrected for extinction) as a function of metallicity
for massive M31 globular clusters (points) and SSP model predictions (thick lines).
Thin lines in the top two panels are fits of $V-K_s$ against metallicity for Milky Way clusters from \citet{cohen07}
and $V-[3.6]$ against metallicity for GCs in NGC~5128 and NGC~4594 from \citet{spitler08}.
Symbols as in Figure~\ref{fig:col_met}; error bars on the data points do not include uncertainties
due to reddening correction, which could be up to 0.3~mag. The very blue cluster at
${\rm [M/H]}=-0.6$ is B037, a cluster with a large ($E(B-V)\gtrsim1.3$) and highly uncertain reddening \citep{caldwell11,barmby02a}.
\label{fig:opt_col_met}}
\end{figure}

\clearpage

Figure~\ref{fig:opt_col_met} shows the $V-K_s$, $V-[3.6]$ and $V-[4.5]$
colors as a function of metallicity for M31 GCs.  Overplotted are the relations predicted by
the GALEV, FSPS, and Padova models (ages 2 and 12 Gyr), as well as fits to the
color-metallicity relationships for other GC systems: $V-K_s$ for Milky Way clusters from \citet{cohen07}
and $V-[3.6]$ from \citet{spitler08}.
In all three model families, older models are redder at a given metallicity than younger models. 
There is a significant spread in predicted color with metallicity: the GALEV models are almost
always redder than the Padova models at a given metallicity, with the FSPS models intermediate.
The Figure shows that the models do not predict a strictly linear color-metallicity relation
for any of the colors; particularly for $V-[3.6]$ and $V-[4.5]$,  all of the models show a slight change in
slope at $[M/H]=-0.7$. The empirical fit in $V-K_s$ captures the model behavior reasonably well, 
although as noted by \citet{conroy10}, the models are redder than the data at high metallicities. 
In $V-[3.6]$, the fit matches the models well at high metallicities but is bluer than the
models at low metallicities. \citet{spitler08} found the same result, although interestingly they were comparing to a
completely different set of models. 

The cluster data points plotted in Figure~\ref{fig:opt_col_met}
are bluer than the models except at the lowest metallicities.
A problem with the photometry does not provide an immediate explanation.
Since our large-aperture IR magnitudes could slightly overestimate the
clusters' brightness (because of contamination by nearby objects), our
photometry seems more prone to make the clusters too red, rather than too blue.
A number of other systematic effects could come into play:
\begin{enumerate}

\item One possibility is that our method of using
the average $\alpha$-element abundances to compute ${\rm [M/H]}$ is too simplistic,
although the ${\rm [M/H]}$ values would have to change drastically to make the
data points compatible with the models.

\item
Another possibility is that  the clusters are much younger than expected and should
therefore be compared to younger models. We think this is unlikely:
deriving ages from mid-IR model predictions and colors alone 
(clearly a risky idea given the limited age sensitivity shown in Figure~\ref{fig:mod_age})
would suggest ages  younger than 2~Gyr.
Such young ages would have a substantial effect on optical and near-infrared colors, and
there are no suggestions from recent spectroscopic or SED-fitting studies
of M31 clusters \citep[e.g.][]{rey07, caldwell09, wang10} that the brightest clusters
are all so young. 

\item 
Reddening corrections could play a role: the metal-rich clusters have a 
higher average reddening than the metal-poor clusters, so  if the $E(B-V)$ estimates
of \citet{caldwell11} were too high, the corrected colors of 
metal-rich clusters would be more strongly affected.
As an example, we found that reducing the reddening values of all clusters by 20\% 
brought the data points and models into better agreement over the entire metallicity range.%
\footnote{The reddening law itself is unlikely to be the problem, since almost
all of the correction to the $V-$infrared colors is in $A_V$ and there is
no strong evidence that M31 has a different value of $R_V$.}

\item All three model families could have some common flaw in the spectral libraries
that causes them to predict too much infrared flux, or too little $V$-band.
Some SSP models are known to have problems reproducing globular cluster colors
in the infrared \citep[Figure 6]{conroy10}, so a problem with the less-well-calibrated
IRAC bands seems plausible.

\item Finally, the results of  \citet{strader09} on the mass-to-light ratios
for M31 clusters suggest an interesting alternative. Those authors 
found $M/L$ to be lower for metal-rich M31 clusters than for metal-poor clusters,
contrary to predictions from SSP models. While younger populations
are expected to have lower $M/L$, \citeauthor{strader09}
 argued against the metal-rich M31  clusters being even
intermediate-age, and concluded that variations in the cluster mass function, either initially or
through dynamical evolution, were more likely explanations.
A flatter mass function, required to match the mass-to-light ratios,  should qualitatively also
result in bluer predicted colors: fewer low-mass stars would mean less $V$-band light (dominated
by the main sequence), and more high-mass stars should mean more infrared light (dominated by giants).
The different initial mass functions available in the SSP models do not result in large
differences in predicted colors, but perhaps more drastic  alterations to the mass function
would change colors more dramatically. 
\end{enumerate}
We consider the last three of these options to be more plausible explanations of the
offset. Distinguishing between them requires better-quality data than are
currently available.

If we assume that the model-data offset in $V-$infrared colors is due to a systematic
problem in the models, we can still use our data to
explore the metallicity sensitivity of these colors.  We performed 
least-squares fits of  color as a function of metallicity, weighted by the color uncertainty. 
For the  clusters binned in metallicity, we found 
slopes in the range  0.42--0.46 for all three colors, very close to the value of 0.409 
derived by \citet{cohen07} for $V-K_s$ in Milky Way
globular clusters. For the full, unbinned M31 sample we derive a slope of 0.39 for $V-K_s$.
For this sample we derive somewhat steeper slopes of 0.68 for  $V-[3.6]$ and 0.59 for $V-[4.5]$,
still shallower than the slope of 0.90 derived by \citet{spitler08}, although those authors were 
fitting the reverse function (metallicity as a function of color).
The steeper slopes for the unbinned data agree better with the general trend of the models.
Of course, there is no guarantee that a linear fit is the correct representation of the
actual physical situation \citep[and see][]{yoon06}, but the scatter visible in Figure~\ref{fig:opt_col_met} 
indicates that  a more complicated fit is not likely to be meaningful.

The slopes of the linear fits indicate that $V-{\rm IRAC}$ colors could be a useful metallicity indicator 
for simple stellar populations. The low infrared background seen by \textit{Spitzer} means that 
IRAC observations to comparable depth can be made much more quickly than ground-based
near-infrared observations; however compared to $V-K_s$, $V-{\rm IRAC}$ colors have the disadvantage of being 
somewhat more sensitive to reddening uncertainties and having (typically) poorer spatial resolution.
Echoing \citet{spitler08}, we caution that these colors are not well-calibrated
for super-solar or very low ($[M/H]\lesssim -1.9$) metallicities. Using the M31 clusters to
calibrate the relation would require careful matching of apertures in photometric bands differing
by almost a factor of 10 in wavelength \citep[see][for a description of the difficulties involved in
matching just between visible and near-infrared]{cohen07} and is not a trivial task even for
these bright clusters. Using more distant, unresolved clusters to calibrate the visible-to-IRAC color/metallicity
relation might be a better approach.

\section{Summary}
\label{sec:summ}

We have presented mid-infrared photometry of massive globular clusters in M31, 
in order to test the  color predictions of population synthesis models.
To our knowledge, this is the first such test of the models in all four IRAC bands.
The three different model families discussed here differ somewhat in their predicted
colors and in the systematic effects of age and metallicity.  As might be expected, 
models including circumstellar dust emission predict redder colors, particularly at  longer wavelengths,
and more color variability with age, particularly for ages  $\lesssim 8$~Gyr.
Except for the $[5.8]-[8.0]$ color, the exact choice of circumstellar dust prescription
does not make a substantial difference.
The present level of observational precision does not permit distinguishing
between models;  the considerable observational uncertainties, especially at
the longer wavelengths, are due to 
the bright background from the galaxy and the falling cluster spectral energy distribution.
Within these uncertainties, the data and models are mostly in agreement. We find
a small but not statistically significant offset between models and data in the $[4.5]-[5.8]$ color;
if real, this could be an indication of emission from circumstellar dust in some clusters.
Finally, we find that visible-to-IRAC colors are bluer than model
predictions, for reasons which are unclear. These colors could still be feasible
metallicity indicators for simple stellar populations but further work is required to
establish accurate calibrations.

Facilities: \facility{Spitzer (IRAC)}, \facility{FLWO:2MASS ()}

\acknowledgments
We thank the anonymous referee for helpful suggestions.
This work is based on observations made with the \textit{Spitzer Space Telescope},
which is operated by the Jet Propulsion Laboratory, California Institute of 
Technology under NASA contract 1407. 
This publication makes use of data products from the Two Micron All Sky Survey, which is 
a joint project of the University of Massachusetts and the Infrared Processing and Analysis 
Center/California Institute of Technology, funded by the National Aeronautics and Space 
Administration and the National Science Foundation.
The authors acknowledge final support from a Discovery Grant (PB) and 
an Undergraduate Summer Research Award (FFJ) from the Natural Sciences
and Engineering Research Council of Canada, and from an Ontario Early
Researcher Award to PB.

\clearpage

\bibliographystyle{aas}

\begin{thebibliography}{}

\bibitem[{{Aringer} {et~al.}(2009)}]{aringer09}
{Aringer}, B., {Girardi}, L., {Nowotny}, W., {Marigo}, P., \& {Lederer}, M.~T.
  2009, \aap, 503, 913
  
\bibitem[{{Barmby} {et~al.}(2009)}]{barmby09}
{Barmby}, P., {Boyer}, M.~L., {Woodward}, C.~E., {Gehrz}, R.~D., {van Loon},
  J.~T., {Fazio}, G.~G., {Marengo}, M., \& {Polomski}, E. 2009, \aj, 137, 207

\bibitem[{{Barmby} \& {Huchra}(2000)}]{barmby00a}
{Barmby}, P. \& {Huchra}, J.~P. 2000, \apjl, 531, L29

\bibitem[{{Barmby} {et~al.}(2007)}]{barmby07}
{Barmby}, P., {McLaughlin}, D.~E., {Harris}, W.~E., {Harris}, G.~L.~H., \&
  {Forbes}, D.~A. 2007, \aj, 133, 2764

\bibitem[{{Barmby} {et~al.}(2002){Barmby}, {Perrett}, \& {Bridges}}]{barmby02a}
{Barmby}, P., {Perrett}, K.~M., \& {Bridges}, T.~J. 2002, \mnras, 329, 461

\bibitem[{{Barmby} {et~al.}(2006)}]{barmby06a}
{Barmby}, P. {et~al.} 2006, \apjl, 650, L45

\bibitem[{{Beasley} {et~al.}(2005)}]{beasley05}
{Beasley}, M.~A., {Brodie}, J.~P., {Strader}, J., {Forbes}, D.~A., {Proctor},
  R.~N., {Barmby}, P., \& {Huchra}, J.~P. 2005, \aj, 129, 1412

\bibitem[{{Bell} \& {de Jong}(2001)}]{bell01}
{Bell}, E.~F. \& {de Jong}, R.~S. 2001, \apj, 550, 212

\bibitem[{{Bessell} {et~al.}(1991)}]{bessell91}
{Bessell}, M.~S., {Brett}, J.~M., {Scholz}, M., \& {Wood}, P.~R. 1991, \aaps,
  89, 335

\bibitem[{{Bessell} {et~al.}(1989)}]{bessell89}
{Bessell}, M.~S., {Brett}, J.~M., {Wood}, P.~R., \& {Scholz}, M. 1989, \aaps,
  77, 1

\bibitem[{{Bica} \& {Alloin}(1986)}]{bica86a}
{Bica}, E. \& {Alloin}, D. 1986, \aaps, 66, 171

\bibitem[{{Blanton} \& {Roweis}(2007)}]{blanton07}
{Blanton}, M.~R. \& {Roweis}, S. 2007, \aj, 133, 734

\bibitem[{{Boyer} {et~al.}(2008)}]{boyer08}
{Boyer}, M.~L., {McDonald}, I., {van Loon}, J.~T., {Woodward}, C.~E., {Gehrz},
  R.~D., {Evans}, A., \& {Dupree}, A.~K. 2008, \aj, 135, 1395

\bibitem[{{Boyer} {et~al.}(2010)}]{boyer10}
{Boyer}, M.~L., et al. 2010, \apjl, 711, L99

\bibitem[{{Boyer} {et~al.}(2006)}]{boyer06}
{Boyer}, M.~L., {Woodward}, C.~E., {van Loon}, J.~T., {Gordon}, K.~D., {Evans},
  A., {Gehrz}, R.~D., {Helton}, L.~A., \& {Polomski}, E.~F. 2006, \aj, 132,
  1415

\bibitem[{{Bragaglia}(2010)}]{bragaglia10}
{Bragaglia}, A. 2010, in IAU Symposium, Vol. 268, IAU Symposium, ed.
  {C.~Charbonnel, M.~Tosi, F.~Primas, \& C.~Chiappini}, 119--128

\bibitem[{{Bressan} {et~al.}(1998){Bressan}, {Granato}, \& {Silva}}]{bressan98}
{Bressan}, A., {Granato}, G.~L., \& {Silva}, L. 1998, \aap, 332, 135

\bibitem[{{Brodie} \& {Strader}(2006)}]{brodie06}
{Brodie}, J.~P. \& {Strader}, J. 2006, \araa, 44, 193

\bibitem[{{Brown} {et~al.}(2004)}]{brown04}
{Brown}, T.~M., {Ferguson}, H.~C., {Smith}, E., {Kimble}, R.~A., {Sweigart},
  A.~V., {Renzini}, A., {Rich}, R.~M., \& {VandenBerg}, D.~A. 2004, \apjl, 613,
  L125

\bibitem[{{Bruzual} \& {Charlot}(2003)}]{bruzual03}
{Bruzual}, G. \& {Charlot}, S. 2003, \mnras, 344, 1000

\bibitem[{{Buzzoni}(1989)}]{buzzoni89}
{Buzzoni}, A. 1989, \apjs, 71, 817

\bibitem[{{Buzzoni}(2005)}]{buzzoni05}
---. 2005, ArXiv Astrophysics e-prints astro-ph/0509602

\bibitem[{{Caldwell} {et~al.}(2009)}]{caldwell09}
{Caldwell}, N., {Harding}, P., {Morrison}, H., {Rose}, J.~A., {Schiavon}, R.,
  \& {Kriessler}, J. 2009, \aj, 137, 94

\bibitem[{{Caldwell} {et~al.}(2011)}]{caldwell11}
{Caldwell}, N., {Schiavon}, R., {Morrison}, H., {Rose}, J.~A., \& {Harding}, P.
  2011, \aj, 141, 61

\bibitem[{{Cambr{\'e}sy} {et~al.}(2002)}]{cambresy02}
{Cambr{\'e}sy}, L., {Beichman}, C.~A., {Jarrett}, T.~H., \& {Cutri}, R.~M.
  2002, \aj, 123, 2559

\bibitem[{{Castelli} \& {Kurucz}(2003)}]{castelli03}
{Castelli}, F. \& {Kurucz}, R.~L. 2003, in IAU Symposium, Vol. 210, Modelling
  of Stellar Atmospheres, ed. {N.~Piskunov, W.~W.~Weiss, \& D.~F.~Gray}, 20

\bibitem[{{Cervi{\~n}o} \& {Luridiana}(2004)}]{cervino04}
{Cervi{\~n}o}, M. \& {Luridiana}, V. 2004, \aap, 413, 145

\bibitem[{{Charlot} {et~al.}(1996){Charlot}, {Worthey}, \&
  {Bressan}}]{charlot96}
{Charlot}, S., {Worthey}, G., \& {Bressan}, A. 1996, \apj, 457, 625

\bibitem[{{Clemens} {et~al.}(2009)}]{clemens09}
{Clemens}, M.~S., {Bressan}, A., {Panuzzo}, P., {Rampazzo}, R., {Silva}, L.,
  {Buson}, L., \& {Granato}, G.~L. 2009, \mnras, 392, 982

\bibitem[{{Cohen}(2006)}]{cohen06}
{Cohen}, J. 2006, \apjl, 653, L21

\bibitem[{{Cohen} {et~al.}(2007)}]{cohen07}
{Cohen}, J.~G., {Hsieh}, S., {Metchev}, S., {Djorgovski}, S.~G., \& {Malkan},
  M. 2007, \aj, 133, 99

\bibitem[{{Conroy} \& {Gunn}(2010)}]{conroy10}
{Conroy}, C. \& {Gunn}, J.~E. 2010, \apj, 712, 833

\bibitem[{{Conroy} {et~al.}(2009){Conroy}, {Gunn}, \& {White}}]{conroy09}
{Conroy}, C., {Gunn}, J.~E., \& {White}, M. 2009, \apj, 699, 486

\bibitem[{{Decin} \& {Eriksson}(2007)}]{decin07}
{Decin}, L. \& {Eriksson}, K. 2007, \aap, 472, 1041

\bibitem[{{Flaherty} {et~al.}(2007)}]{flaherty07}
{Flaherty}, K.~M., {Pipher}, J.~L., {Megeath}, S.~T., {Winston}, E.~M.,
  {Gutermuth}, R.~A., {Muzerolle}, J., {Allen}, L.~E., \& {Fazio}, G.~G. 2007,
  \apj, 663, 1069

\bibitem[{{Fluks} {et~al.}(1994)}]{fluks94}
{Fluks}, M.~A., {Plez}, B., {The}, P.~S., {de Winter}, D., {Westerlund}, B.~E.,
  \& {Steenman}, H.~C. 1994, \aaps, 105, 311


\bibitem[{{Forbes} {et~al.}(2000)}]{forbes00}
{Forbes}, D.~A., {Masters}, K.~L., {Minniti}, D., \& {Barmby}, P. 2000, \aap,
  358, 471

\bibitem[{{Fouesneau} \& {Lan{\c c}on}(2010)}]{fouesneau10}
{Fouesneau}, M. \& {Lan{\c c}on}, A. 2010, \aap, 521, A22

\bibitem[{{Galleti} {et~al.}(2004)}]{rbc04}
{Galleti}, S., {Federici}, L., {Bellazzini}, M., {Fusi Pecci}, F., \&
  {Macrina}, S. 2004, \aap, 416, 917

\bibitem[{{Gao} {et~al.}(2009){Gao}, {Jiang}, \& {Li}}]{gao09}
{Gao}, J., {Jiang}, B.~W., \& {Li}, A. 2009, \apj, 707, 89

\bibitem[{{Girardi} {et~al.}(2010)}]{girardi10}
{Girardi}, L., et al.
  2010, \apj, 724, 1030

\bibitem[{{Gordon} {et~al.}(2008)}]{gordon08}
{Gordon}, K.~D., {Engelbracht}, C.~W., {Rieke}, G.~H., {Misselt}, K.~A.,
  {Smith}, J.-D.~T., \& {Kennicutt}, Jr., R.~C. 2008, \apj, 682, 336

\bibitem[{{Gordon} {et~al.}(2005)}]{gordon05}
{Gordon}, K.~D. {et~al.} 2005, \pasp, 117, 503

\bibitem[{{Groenewegen}(2006)}]{groenewegen06}
{Groenewegen}, M.~A.~T. 2006, \aap, 448, 181

\bibitem[{{Gutermuth} {et~al.}(2009)}]{gutermuth09}
{Gutermuth}, R.~A., {Megeath}, S.~T., {Myers}, P.~C., {Allen}, L.~E., {Pipher},
  J.~L., \& {Fazio}, G.~G. 2009, \apjs, 184, 18

\bibitem[{{Harris}(2010)}]{harris10}
{Harris}, W.~E. 2010, ArXiv e-prints, 1012.3224

\bibitem[{{Hodge} {et~al.}(2004){Hodge}, {Kraemer}, {Price}, \&
  {Walker}}]{hodge04}
{Hodge}, T.~M., {Kraemer}, K.~E., {Price}, S.~D., \& {Walker}, H.~J. 2004,
  \apjs, 151, 299

\bibitem[{{Indebetouw} {et~al.}(2005)}]{indebetouw05}
{Indebetouw}, R., et al. 2005, \apj, 619, 931

\bibitem[{{Kotulla} {et~al.}(2009)}]{kotulla09}
{Kotulla}, R., {Fritze}, U., {Weilbacher}, P., \& {Anders}, P. 2009, \mnras,
  396, 462

\bibitem[{{Kroupa}(2001)}]{kroupa01}
{Kroupa}, P. 2001, \mnras, 322, 231

\bibitem[{{Kurucz}(1992)}]{kurucz92}
{Kurucz}, R.~L. 1992, in IAU Symposium, Vol. 149, The Stellar Populations of
  Galaxies, ed. {B.~Barbuy \& A.~Renzini}, 225

\bibitem[{{Lan{\c c}on} \& {Mouhcine}(2002)}]{lancon02}
{Lan{\c c}on}, A. \& {Mouhcine}, M. 2002, \aap, 393, 167

\bibitem[{{Lejeune} {et~al.}(1998){Lejeune}, {Cuisinier}, \&
  {Buser}}]{lejeune98}
{Lejeune}, T., {Cuisinier}, F., \& {Buser}, R. 1998, \aaps, 130, 65

\bibitem[{{Loidl} {et~al.}(2001){Loidl}, {Lan{\c c}on}, \&
  {J{\o}rgensen}}]{loidl01}
{Loidl}, R., {Lan{\c c}on}, A., \& {J{\o}rgensen}, U.~G. 2001, \aap, 371, 1065

\bibitem[{{Maraston}(2005)}]{maraston05}
{Maraston}, C. 2005, \mnras, 362, 799

\bibitem[{{Marigo} {et~al.}(2008)}]{marigo08}
{Marigo}, P., {Girardi}, L., {Bressan}, A., {Groenewegen}, M.~A.~T., {Silva},
  L., \& {Granato}, G.~L. 2008, \aap, 482, 883

\bibitem[{{Matsunaga} {et~al.}(2008)}]{matsunaga08}
{Matsunaga}, N., {Mito}, H., {Nakada}, Y., {Fukushi}, H., {Tanab{\'e}}, T.,
  {Ita}, Y., {Izumiura}, H., {Matsuura}, M., {Ueta}, T., \& {Yamamura}, I.
  2008, \pasj, 60, 415

\bibitem[{{McConnachie} {et~al.}(2005)}]{mcconnachie05}
{McConnachie}, A.~W., {Irwin}, M.~J., {Ferguson}, A.~M.~N., {Ibata}, R.~A.,
  {Lewis}, G.~F., \& {Tanvir}, N. 2005, \mnras, 356, 979

\bibitem[{{McLaughlin}(2000)}]{mclaughlin00}
{McLaughlin}, D.~E. 2000, \apj, 539, 618

\bibitem[{{McLaughlin} \& {van der Marel}(2005)}]{mclaughlin05}
{McLaughlin}, D.~E. \& {van der Marel}, R.~P. 2005, \apjs, 161, 304

\bibitem[{{Mould} {et~al.}(2008)}]{mould08}
{Mould}, J., {Barmby}, P., {Gordon}, K., {Willner}, S.~P., {Ashby}, M.~L.~N.,
  {Gehrz}, R.~D., {Humphreys}, R., \& {Woodward}, C.~E. 2008, \apj, 687, 230

\bibitem[{{Nantais} {et~al.}(2006)}]{nantais06}
{Nantais}, J.~B., {Huchra}, J.~P., {Barmby}, P., {Olsen}, K.~A.~G., \&
  {Jarrett}, T.~H. 2006, \aj, 131, 1416

\bibitem[{{Nishiyama} {et~al.}(2009)}]{nishiyama09}
{Nishiyama}, S., {Tamura}, M., {Hatano}, H., {Kato}, D., {Tanab{\'e}}, T.,
  {Sugitani}, K., \& {Nagata}, T. 2009, \apj, 696, 1407

\bibitem[{{Origlia} {et~al.}(2007)}]{origlia07}
{Origlia}, L., {Rood}, R.~T., {Fabbri}, S., {Ferraro}, F.~R., {Fusi Pecci}, F.,
  \& {Rich}, R.~M. 2007, \apjl, 667, L85

\bibitem[{{Origlia} {et~al.}(2010)}]{origlia10}
{Origlia}, L., {Rood}, R.~T., {Fabbri}, S., {Ferraro}, F.~R., {Fusi Pecci}, F.,
  {Rich}, R.~M., \& {Dalessandro}, E. 2010, \apj, 718, 522

\bibitem[{{Pahre} {et~al.}(2004)}]{pahre04}
{Pahre}, M.~A., {Ashby}, M.~L.~N., {Fazio}, G.~G., \& {Willner}, S.~P. 2004,
  \apjs, 154, 229

\bibitem[{{Peacock} {et~al.}(2010)}]{peacock10a}
{Peacock}, M.~B., {Maccarone}, T.~J., {Knigge}, C., {Kundu}, A., {Waters},
  C.~Z., {Zepf}, S.~E., \& {Zurek}, D.~R. 2010, \mnras, 402, 803

\bibitem[{{Peacock} {et~al.}(2011)}]{peacock11}
{Peacock}, M.~B., {Zepf}, S.~E., {Maccarone}, T.~J., \& {Kundu}, A. 2011, ApJ,
  in press, arxiv:1105.3365

\bibitem[{{Perina} {et~al.}(2011)}]{perina11}
{Perina}, S., {Galleti}, S., {Fusi Pecci}, F., {Bellazzini}, M., {Federici},
  L., \& {Buzzoni}, A. 2011, \aap, 531, A155

\bibitem[{{Pessev} {et~al.}(2010){Pessev}, {Goudfrooij}, {Puzia}, \&
  {Chandar}}]{pessev10}
{Pessev}, P., {Goudfrooij}, P., {Puzia}, T., \& {Chandar}, R. 2010, in Bulletin
  of the American Astronomical Society, Vol.~42, American Astronomical Society
  Meeting 215, 425.24

\bibitem[{{Pessev} {et~al.}(2006)}]{pessev06}
{Pessev}, P.~M., {Goudfrooij}, P., {Puzia}, T.~H., \& {Chandar}, R. 2006, \aj,
  132, 781

\bibitem[{{Pessev} {et~al.}(2008)}]{pessev08}
---. 2008, \mnras, 385, 1535

\bibitem[{{Renzini}(1998)}]{renzini98}
{Renzini}, A. 1998, \aj, 115, 2459

\bibitem[{{Rey} {et~al.}(2007)}]{rey07}
{Rey}, S.-C. {et~al.} 2007, \apjs, 173, 643

\bibitem[{{Riffel} {et~al.}(2010)}]{riffel10}
{Riffel}, R., {Ruschel-Dutra}, D., {Pastoriza}, M.~G.,
  {Rodr{\'{\i}}guez-Ardila}, A., {Santos}, Jr., J.~F.~C., {Bonatto}, C.~J., \&
  {Ducati}, J.~R. 2010, \mnras, 1542

\bibitem[{{Rom{\'a}n-Z{\'u}{\~n}iga} {et~al.}(2007)}]{roman-zuniga07}
{Rom{\'a}n-Z{\'u}{\~n}iga}, C.~G., {Lada}, C.~J., {Muench}, A., \& {Alves},
  J.~F. 2007, \apj, 664, 357

\bibitem[{{Salpeter}(1955)}]{salpeter55}
{Salpeter}, E.~E. 1955, \apj, 121, 161

\bibitem[{{Sarajedini} \& {Mancone}(2007)}]{sarajedini07}
{Sarajedini}, A. \& {Mancone}, C.~L. 2007, \aj, 134, 447

\bibitem[{{Schiavon}(2007)}]{schiavon07}
{Schiavon}, R.~P. 2007, \apjs, 171, 146

\bibitem[{{Sheth} {et~al.}(2010){Sheth}}]{sheth10}
{Sheth}, K., et al. 2010, \pasp, 122, 1397

\bibitem[{{Skrutskie} {et~al.}(2006)}]{skrutskie06}
{Skrutskie}, M.~F. {et~al.} 2006, \aj, 131, 1163

\bibitem[{{Sloan} {et~al.}(2003){Sloan}, {Kraemer}, {Price}, \&
  {Shipman}}]{sloan03}
{Sloan}, G.~C., {Kraemer}, K.~E., {Price}, S.~D., \& {Shipman}, R.~F. 2003,
  \apjs, 147, 379

\bibitem[{{Spitler} {et~al.}(2008){Spitler}, {Forbes}, \&
  {Beasley}}]{spitler08}
{Spitler}, L.~R., {Forbes}, D.~A., \& {Beasley}, M.~A. 2008, \mnras, 389, 1150

\bibitem[{{Spitzer Science Center}(2010)}]{iracihb}
{Spitzer Science Center}. 2010, {IRAC Instrument Handbook, v1.0} (Pasadena, CA: SSC)

\bibitem[{{Strader} {et~al.}(2009)}]{strader09}
{Strader}, J., {Smith}, G.~H., {Larsen}, S., {Brodie}, J.~P., \& {Huchra},
  J.~P. 2009, \aj, 138, 547

\bibitem[{{Thomas} {et~al.}(2003){Thomas}, {Maraston}, \& {Bender}}]{thomas03}
{Thomas}, D., {Maraston}, C., \& {Bender}, R. 2003, \mnras, 339, 897

\bibitem[{{Tinsley} \& {Gunn}(1976)}]{tinsley76}
{Tinsley}, B.~M. \& {Gunn}, J.~E. 1976, \apj, 203, 52

\bibitem[{{Vandenbussche} {et~al.}(2002)}]{vandenbussche02}
{Vandenbussche}, B., et al.
  2002, \aap, 390, 1033

\bibitem[{{Wang} {et~al.}(2010)}]{wang10}
{Wang}, S., {Fan}, Z., {Ma}, J., {de Grijs}, R., \& {Zhou}, X. 2010, \aj, 139,
  1438

\bibitem[{{Westera} {et~al.}(2002)}]{westera02}
{Westera}, P., {Lejeune}, T., {Buser}, R., {Cuisinier}, F., \& {Bruzual}, G.
  2002, \aap, 381, 524

\bibitem[{{Yoon} {et~al.}(2006){Yoon}, {Yi}, \& {Lee}}]{yoon06}
{Yoon}, S.-J., {Yi}, S.~K., \& {Lee}, Y.-W. 2006, Science, 311, 1129

\bibitem[{{Zasowski} {et~al.}(2009)}]{zasowski09}
{Zasowski}, G., et al. 2009, \apj, 707, 510

\end{thebibliography}

\begin{deluxetable}{llllllllll}
\tabletypesize{\tiny}
\tablecaption{IRAC photometry for massive M31 globular clusters\label{tab:gcphot}}
\tablewidth{0pt}
\tablehead{
\colhead{Name\tablenotemark{a}}   & \multicolumn{4}{c}{3\farcs66 radius aperture}& \multicolumn{3}{c}{12\farcs2 radius aperture} & 
\colhead{${\rm [Fe/H]}$\tablenotemark{b}} & 
\colhead{$E(B-V)$\tablenotemark{b}} \\

\colhead{}  & \colhead{[3.6]} & \colhead{[4.5]} & \colhead{[5.8]} & \colhead{[8.0]} &  \colhead{$K_s$} & \colhead{[3.6]} & \colhead{[4.5]} & 
\colhead{} & \colhead{}
}
\startdata
B006   & $12.90 \pm 0.01 $ & $ 12.97 \pm 0.01 $ & $12.89 \pm 0.02 $ & $12.81 \pm 0.02 $ & $ 12.43 \pm 0.03 $ & $ 12.27 \pm 0.01 $ & $12.32 \pm 0.01$ & $-0.5$ & 0.17\\
B012   & $12.99 \pm 0.01 $ & $ 13.00 \pm 0.01 $ & $12.91 \pm 0.01 $ & $12.89 \pm 0.01 $ & $ 12.70 \pm 0.04 $ & $ 12.67 \pm 0.01 $ & $12.60 \pm 0.01$ & $-1.7$ & 0.17\\
B017   & $12.85 \pm 0.01 $ & $ 12.80 \pm 0.01 $ & $12.78 \pm 0.01 $ & $12.63 \pm 0.02 $ & $ 12.44 \pm 0.04 $ & $ 12.36 \pm 0.01 $ & $12.30 \pm 0.01$ & $-0.8$ & 0.47\\
B019   & $12.17 \pm 0.01 $ & $ 12.28 \pm 0.01 $ & $12.21 \pm 0.01 $ & $12.13 \pm 0.02 $ & $ 11.93 \pm 0.03 $ & $ 11.70 \pm 0.01 $ & $11.79 \pm 0.01$ & $-0.8$ & 0.22\\
B023   & $11.03 \pm 0.01 $ & $ 11.11 \pm 0.01 $ & $10.95 \pm 0.01 $ & $10.91 \pm 0.01 $ & $ 10.73 \pm 0.01 $ & $ 10.59 \pm 0.01 $ & $10.67 \pm 0.01$ & $-0.7$ & 0.42\\
B027   & $13.23 \pm 0.01 $ & $ 13.26 \pm 0.01 $ & $13.29 \pm 0.03 $ & $13.28 \pm 0.04 $ & $ 13.16 \pm 0.12 $ & $ 12.85 \pm 0.04 $ & $12.89 \pm 0.03$ & $-1.3$ & 0.19\\
B034   & $12.74 \pm 0.01 $ & $ 12.77 \pm 0.01 $ & $12.67 \pm 0.02 $ & $12.67 \pm 0.03 $ & $ 12.37 \pm 0.06 $ & $ 12.09 \pm 0.01 $ & $12.12 \pm 0.01$ & $-0.6$ & 0.16\\
B037   & $11.20 \pm 0.01 $ & $ 11.25 \pm 0.01 $ & $11.12 \pm 0.01 $ & $11.02 \pm 0.01 $ & $ 10.94 \pm 0.02 $ & $ 10.74 \pm 0.01 $ & $10.79 \pm 0.01$ & $-0.8$ & 1.61\\
B051   & $13.07 \pm 0.01 $ & $ 13.14 \pm 0.01 $ & $13.05 \pm 0.02 $ & $13.12 \pm 0.04 $ & $ 12.49 \pm 0.05 $ & $ 12.35 \pm 0.01 $ & $12.38 \pm 0.01$ & $-0.8$ & 0.38\\
B055   & $12.86 \pm 0.02 $ & $ 12.85 \pm 0.02 $ & $12.57 \pm 0.03 $ & $12.05 \pm 0.03 $ & $ 12.34 \pm 0.09 $ & $ 11.98 \pm 0.02 $ & $11.90 \pm 0.02$ & $-0.1$ & 0.54\\
B058   & $12.76 \pm 0.01 $ & $ 12.71 \pm 0.01 $ & $12.76 \pm 0.02 $ & $12.65 \pm 0.04 $ & $ 12.33 \pm 0.05 $ & $ 12.15 \pm 0.01 $ & $12.14 \pm 0.01$ & $-1.1$ & 0.15\\
B061   & $13.16 \pm 0.01 $ & $ 13.20 \pm 0.01 $ & $13.10 \pm 0.02 $ & $13.04 \pm 0.03 $ & $ 12.90 \pm 0.08 $ & $ 12.58 \pm 0.02 $ & $12.61 \pm 0.02$ & $-0.7$ & 0.49\\
B063   & $12.18 \pm 0.01 $ & $ 12.26 \pm 0.01 $ & $12.10 \pm 0.01 $ & $11.98 \pm 0.01 $ & $ 11.92 \pm 0.03 $ & $ 11.75 \pm 0.01 $ & $11.82 \pm 0.01$ & $-0.8$ & 0.49\\
B068   & $12.91 \pm 0.01 $ & $ 12.94 \pm 0.01 $ & $12.84 \pm 0.02 $ & $12.76 \pm 0.04 $ & $ 12.56 \pm 0.13 $ & $ 12.23 \pm 0.02 $ & $12.20 \pm 0.02$ & $-0.2$ & 0.45\\
B082   & $11.64 \pm 0.01 $ & $ 11.66 \pm 0.01 $ & $11.58 \pm 0.01 $ & $11.50 \pm 0.02 $ & $ 11.38 \pm 0.09 $ & $ 11.19 \pm 0.01 $ & $11.20 \pm 0.01$ & $-0.7$ & 0.94\\
B086   & $12.95 \pm 0.03 $ & $ 13.02 \pm 0.03 $ & $12.76 \pm 0.03 $ & $12.64 \pm 0.05 $ & $ 12.63 \pm 0.26 $ & $ 12.47 \pm 0.10 $ & $12.71 \pm 0.10$ & $-1.8$ & 0.13\\
B088   & $12.50 \pm 0.01 $ & $ 12.49 \pm 0.01 $ & $12.47 \pm 0.01 $ & $12.36 \pm 0.02 $ & $ 12.12 \pm 0.04 $ & $ 11.96 \pm 0.01 $ & $11.96 \pm 0.01$ & $-1.8$ & 0.53\\
B091D  & $12.29 \pm 0.03 $ & $ 12.41 \pm 0.03 $ & $12.32 \pm 0.03 $ & $12.20 \pm 0.04 $ &  \nodata\tablenotemark{c} & \nodata & \nodata  & $-0.7$ & 0.24\\
B094   & $12.70 \pm 0.01 $ & $ 12.74 \pm 0.01 $ & $12.65 \pm 0.02 $ & $12.44 \pm 0.04 $ & $ 12.22 \pm 0.05 $ & $ 12.06 \pm 0.01 $ & $12.11 \pm 0.01$ & $-0.4$ & 0.22\\
B096   & $13.19 \pm 0.03 $ & $ 13.28 \pm 0.04 $ & $13.27 \pm 0.04 $ & $13.30 \pm 0.13 $ & $ 12.54 \pm 0.23 $ & $ 12.50 \pm 0.09 $ & $12.46 \pm 0.07$ & $-0.3$ & 0.63\\
B103   & $12.20 \pm 0.03 $ & $ 12.27 \pm 0.02 $ & $12.18 \pm 0.03 $ & $12.03 \pm 0.06 $ & $ 11.46 \pm 0.17 $ & $ 11.30 \pm 0.01 $ & $11.43 \pm 0.01$ & $-0.5$ & 0.34\\
B107   & $13.25 \pm 0.04 $ & $ 13.07 \pm 0.03 $ & $13.05 \pm 0.06 $ & $12.92 \pm 0.13 $ & $ 12.59 \pm 0.29 $ & $ 12.32 \pm 0.10 $ & $12.16 \pm 0.08$ & $-1.0$ & 0.22\\
B116   & $12.84 \pm 0.01 $ & $ 12.85 \pm 0.01 $ & $12.76 \pm 0.02 $ & $12.72 \pm 0.04 $ & $ 12.35 \pm 0.04 $ & $ 12.17 \pm 0.01 $ & $12.19 \pm 0.01$ & $-0.6$ & 0.72\\
B124   & $11.70 \pm 0.10 $ & $ 11.72 \pm 0.09 $ & $11.47 \pm 0.10 $ & $11.41 \pm 0.11 $ &  \nodata\tablenotemark{d} & \nodata & \nodata  & $-0.5$ & 0.17\\
B131   & $12.81 \pm 0.09 $ & $ 12.79 \pm 0.08 $ & $12.60 \pm 0.09 $ & $12.34 \pm 0.11 $ &  \nodata\tablenotemark{d} & \nodata & \nodata  & $-0.7$ & 0.18\\
B135   & $13.33 \pm 0.01 $ & $ 13.34 \pm 0.01 $ & $13.22 \pm 0.02 $ & $13.16 \pm 0.04 $ & $ 13.05 \pm 0.10 $ & $ 12.98 \pm 0.03 $ & $12.95 \pm 0.03$ & $-1.8$ & 0.28\\
B143   & $13.03 \pm 0.05 $ & $ 13.14 \pm 0.05 $ & $12.94 \pm 0.05 $ & $12.75 \pm 0.06 $ & $ 12.71 \pm 0.41 $ & $ 12.44 \pm 0.01 $ & $12.44 \pm 0.01$ & $-0.1$ & 0.34\\
B148   & $13.52 \pm 0.08 $ & $ 13.51 \pm 0.08 $ & $13.49 \pm 0.11 $ & $13.04 \pm 0.09 $ & $ 12.78 \pm 0.46 $ & $ 12.27 \pm 0.04 $ & $12.34 \pm 0.04$ & $-1.1$ & 0.28\\
B151   & $11.57 \pm 0.01 $ & $ 11.65 \pm 0.01 $ & $11.46 \pm 0.01 $ & $11.39 \pm 0.02 $ & $ 11.20 \pm 0.05 $ & $ 11.03 \pm 0.01 $ & $11.14 \pm 0.01$ & $-0.6$ & 0.53\\
B163   & $11.90 \pm 0.01 $ & $ 11.97 \pm 0.01 $ & $11.86 \pm 0.01 $ & $11.68 \pm 0.01 $ & $ 11.50 \pm 0.03 $ & $ 11.32 \pm 0.01 $ & $11.40 \pm 0.01$ & $-0.1$ & 0.21\\
B171   & $12.35 \pm 0.01 $ & $ 12.35 \pm 0.01 $ & $12.25 \pm 0.01 $ & $12.02 \pm 0.02 $ & $ 11.90 \pm 0.07 $ & $ 11.79 \pm 0.03 $ & $11.79 \pm 0.02$ & $-0.3$ & 0.19\\
B174   & $12.79 \pm 0.01 $ & $ 12.86 \pm 0.01 $ & $12.79 \pm 0.02 $ & $12.76 \pm 0.03 $ & $ 12.33 \pm 0.05 $ & $ 12.18 \pm 0.01 $ & $12.26 \pm 0.02$ & $-1.0$ & 0.28\\
B178   & $12.90 \pm 0.01 $ & $ 12.91 \pm 0.01 $ & $12.84 \pm 0.02 $ & $12.71 \pm 0.03 $ & $ 12.65 \pm 0.12 $ & $ 12.64 \pm 0.05 $ & $12.54 \pm 0.05$ & $-1.2$ & 0.10\\
B179   & $13.05 \pm 0.02 $ & $ 13.08 \pm 0.02 $ & $13.08 \pm 0.03 $ & $12.86 \pm 0.07 $ & $ 12.67 \pm 0.17 $ & $ 12.53 \pm 0.06 $ & $12.37 \pm 0.05$ & $-1.0$ & 0.10\\
B182   & $12.74 \pm 0.01 $ & $ 12.83 \pm 0.02 $ & $12.71 \pm 0.03 $ & $12.63 \pm 0.06 $ & $ 12.16 \pm 0.04 $ & $ 11.84 \pm 0.01 $ & $11.84 \pm 0.01$ & $-1.0$ & 0.33\\
B185   & $12.84 \pm 0.01 $ & $ 12.88 \pm 0.01 $ & $12.79 \pm 0.02 $ & $12.75 \pm 0.04 $ & $ 12.54 \pm 0.09 $ & $ 12.49 \pm 0.04 $ & $12.61 \pm 0.04$ & $-0.6$ & 0.21\\
B193   & $12.43 \pm 0.01 $ & $ 12.48 \pm 0.01 $ & $12.36 \pm 0.01 $ & $12.16 \pm 0.02 $ & $ 12.03 \pm 0.05 $ & $ 11.77 \pm 0.01 $ & $11.66 \pm 0.01$ & $-0.1$ & 0.23\\
B206   & $12.78 \pm 0.02 $ & $ 12.76 \pm 0.02 $ & $12.75 \pm 0.03 $ & $12.62 \pm 0.03 $ & $ 12.19 \pm 0.06 $ & $ 12.15 \pm 0.02 $ & $12.13 \pm 0.02$ & $-1.1$ & 0.10\\
B212   & $13.42 \pm 0.01 $ & $ 13.45 \pm 0.01 $ & $13.36 \pm 0.02 $ & $13.35 \pm 0.04 $ & $ 12.96 \pm 0.07 $ & $ 12.74 \pm 0.01 $ & $12.83 \pm 0.01$ & $-1.7$ & 0.16\\
B224   & $13.72 \pm 0.03 $ & $ 13.74 \pm 0.03 $ & $13.73 \pm 0.03 $ & $13.74 \pm 0.07 $ & $ 12.95 \pm 0.11 $ & $ 12.78 \pm 0.03 $ & $12.72 \pm 0.03$ & $-1.5$ & 0.15\\
B225   & $11.28 \pm 0.01 $ & $ 11.32 \pm 0.01 $ & $11.06 \pm 0.01 $ & $10.67 \pm 0.02 $ & $ 11.07 \pm 0.02 $ & $ 10.89 \pm 0.01 $ & $10.91 \pm 0.01$ & $-0.5$ & 0.12\\
B232   & $13.54 \pm 0.01 $ & $ 13.54 \pm 0.01 $ & $13.48 \pm 0.02 $ & $13.37 \pm 0.04 $ & $ 13.13 \pm 0.07 $ & $ 12.89 \pm 0.01 $ & $12.88 \pm 0.01$ & $-1.9$ & 0.21\\
B306   & $12.84 \pm 0.01 $ & $ 12.89 \pm 0.01 $ & $12.84 \pm 0.02 $ & $12.77 \pm 0.03 $ & $ 12.44 \pm 0.03 $ & $ 12.23 \pm 0.01 $ & $12.20 \pm 0.01$ & $-1.1$ & 0.57\\
B311   & $13.09 \pm 0.01 $ & $ 13.10 \pm 0.01 $ & $13.05 \pm 0.02 $ & $13.04 \pm 0.02 $ & $ 12.60 \pm 0.05 $ & $ 12.42 \pm 0.01 $ & $12.42 \pm 0.01$ & $-1.9$ & 0.36\\
B312   & $13.04 \pm 0.01 $ & $ 13.13 \pm 0.01 $ & $13.01 \pm 0.02 $ & $13.04 \pm 0.03 $ & $ 12.60 \pm 0.04 $ & $ 12.46 \pm 0.01 $ & $12.59 \pm 0.01$ & $-1.2$ & 0.23\\
B338   & $12.01 \pm 0.01 $ & $ 12.02 \pm 0.01 $ & $11.89 \pm 0.01 $ & $11.87 \pm 0.03 $ & $ 11.53 \pm 0.03 $ & $ 11.52 \pm 0.01 $ & $11.53 \pm 0.01$ & $-1.1$ & 0.15\\
B358   & $13.33 \pm 0.01 $ & $ 13.32 \pm 0.01 $ & $13.21 \pm 0.01 $ & $13.19 \pm 0.02 $ & $ 12.96 \pm 0.12 $ & $ 12.76 \pm 0.03 $ & $12.74 \pm 0.02$ & $-1.9$ & 0.06\\
B373   & $12.68 \pm 0.01 $ & $ 12.72 \pm 0.01 $ & $12.71 \pm 0.02 $ & $12.71 \pm 0.07 $ & $ 12.29 \pm 0.03 $ & $ 12.18 \pm 0.01 $ & $12.20 \pm 0.01$ & $-0.5$ & 0.25\\
B386   & $13.15 \pm 0.01 $ & $ 13.17 \pm 0.01 $ & $13.12 \pm 0.02 $ & $13.09 \pm 0.03 $ & $ 12.72 \pm 0.11 $ & $ 12.44 \pm 0.01 $ & $12.43 \pm 0.01$ & $-1.1$ & 0.18\\
B472   & $12.85 \pm 0.02 $ & $ 12.80 \pm 0.02 $ & $12.73 \pm 0.03 $ & $12.46 \pm 0.04 $ & $ 12.39 \pm 0.10 $ & $ 12.20 \pm 0.03 $ & $12.13 \pm 0.03$ & $-1.0$ & 0.12\\
G001   & $11.34 \pm 0.01 $ & $ 11.37 \pm 0.01 $ & $11.22 \pm 0.01 $ & $11.12 \pm 0.01 $ & $ 10.61 \pm 0.02 $ & $ 10.55 \pm 0.01 $ & $10.57 \pm 0.01$ & $-1.0$ & 0.10\\
MITA140& $12.72 \pm 0.02 $ & $ 12.75 \pm 0.01 $ & $12.68 \pm 0.02 $ & $12.56 \pm 0.05 $ & $ 12.40 \pm 0.13 $ & $ 12.30 \pm 0.05 $ & $12.23 \pm 0.04$ & $-0.2$ & 0.86\\
\enddata
\tablecomments{Values are not corrected for extinction. Measurements in the 5.8 and 8.0~\micron{} bands are made only in the smaller aperture because of the
brighter background at these wavelengths.}
\tablenotetext{a}{Naming convention is that of \citet{rbc04}.}
\tablenotetext{b}{From \citet{caldwell11}.}
\tablenotetext{c}{This cluster is located near a much brighter star; adequate large-aperture photometry was not possible.} 
\tablenotetext{d}{These clusters are located near the bright bulge of M31; adequate large-aperture photometry was not possible.} 
\end{deluxetable}

\end{document}